\documentclass[nofootinbib,a4paper,fleqn,aps,prd,10pt,superscriptaddress,reprint,showkeys,showpacs]{revtex4-1}

\usepackage{times}
\usepackage{amsfonts}
\usepackage{amsmath}
\usepackage{amssymb}
\usepackage{natbib}
\usepackage{graphicx}
\usepackage{enumitem}
\usepackage{xcolor}
\usepackage[colorlinks=true,linkcolor=blue]{hyperref}
\usepackage[outdir=./]{epstopdf}
\usepackage[normalem]{ulem}

\def\aap{A\&A}
\def\aj{AJ}
\def\apj{ApJ}
\def\apjl{ApJL}
\def\apjs{ApJS}

\def\jcap{JCAP}
\def\mnras{MNRAS}
\def\na{New Astronomy}

\begin{document}

\title{{\bf Turnaround radius in $\Lambda$CDM}, and dark matter cosmologies II: the role of dynamical friction
%
%
}

\author{Antonino~\surname{Del Popolo}}%
\affiliation{%
Dipartimento di Fisica e Astronomia, University of Catania, Viale Andrea Doria 6, 95125, Catania, Italy
}
\affiliation{%
Institute of Astronomy, Russian Academy of Sciences, 119017, Pyatnitskaya str., 48 , Moscow 
}
\affiliation{%
INFN sezione di Catania, Via S. Sofia 64, I-95123 Catania, Italy
}
\email[Corresponding author: ]{adelpopolo@oact.inaf.it}

\author{Man Ho~\surname{Chan}}%
\affiliation{Department of Science and Environmental Studies, The Education University of Hong Kong, Tai Po, New Territories, Hong Kong }
\email[]{chanmh@eduhk.hk}



\label{firstpage}

\date{\today}

\begin{abstract}
This paper is an extension of the paper by Del Popolo, Chan, and Mota (2020) to take account the effect of dynamical friction. We show how dynamical friction changes the threshold of collapse, $\delta_c$, and the turn-around radius, $R_t$. We find numerically 
the relationship between the turnaround radius, $R_{\rm t}$, and mass, $M_{\rm t}$, in $\Lambda$CDM, in dark energy scenarios, and in a $f(R)$ modified gravity model. Dynamical friction gives rise to a $R_{\rm t}-M_{\rm t}$ relation differing from that of the standard spherical collapse. In particular, dynamical friction amplifies the effect of shear, and vorticity already studied in Del Popolo, Chan, and Mota (2020). A comparison of the  $R_{\rm t}-M_{\rm t}$ relationship for the $\Lambda$CDM, and those for the dark energy, and modified gravity models shows, that the $R_{\rm t}-M_{\rm t}$ relationship of the $\Lambda$CDM is similar to that of the dark energy models, and small differences are seen when comparing with the $f(R)$ models. 
The effect of shear, rotation, and dynamical friction is particularly evident at galactic scales, giving rise to a difference between the $R_{\rm t}-M_{\rm t}$ relation of the standard spherical collapse of the order of $\simeq 60\%$. Finally, we show how the new values of the $R_{\rm t}-M_{\rm t}$ influence the constraints to the $w$ parameter of the equation of state. 
\end{abstract}



\maketitle

\section{Introduction}


In the past several decades, observations revealed that some missing mass exist in our universe \cite{Sanders}. Many physicists believe that the existence of some unknown particles called cold dark matter (CDM) can account for the missing mass \cite{Profumo}. On the other hand, cosmological observations suggest that the expansion of our universe is accelerating. Many cosmologists propose that the existence of a new kind of energy called dark energy can help explain the accelerating expansion \cite{Li2011}. In standard cosmological model, the amount of dark energy can be represented by the cosmological constant $\Lambda$. This standard cosmological model is now known as the $\Lambda$CDM model.
The $\Lambda$CDM model can give good agreements for observations in large-scale structures \cite{DelPopolo2013,Ade}. However, there are some discrepancies between the predictions of the $\Lambda$CDM model and the observations in small-scale structures. Specifically, the core-cusp problem \cite{deBlok2010}, the missing satellites problem \cite{Moore} and the mass-discrepancy acceleration relation problem \cite{McGaugh1,McGaugh2} are three classical problems challenging the $\Lambda$CDM model. Moreover, currently no compelling particle dark matter signal has been detected directly or indirectly. The current direct-detection and indirect-detection constraints of dark matter have ruled out a large parameter space of potential particle dark matter models \cite{Aprile,Abecrcrombie,Ackermann,Chan2017,Chan2019,Chan3}. Also, the cosmological constant $\Lambda$ suffers from the cosmological constant fine-tuning problem and the cosmic coincidence problem \cite{Weinberg1989,Velten2014}. Therefore, despite some success in the cosmological scale, the $\Lambda$CDM model is still being challenged by many recent studies.

Based on the above problems, some studies propose alternative models for the universe accelerated expansion. Dark energy (DE) effects are generated by additional matter fields (e.g., quintessence \citep{Copeland2006}), or 
that the dynamical effects of dark matter and/or dark energy might originate from modified gravity (MG)
models\citep{Horndeski1974,Milgrom1983,Zwiebach1985,Moffat2006,Nojiri2005,Bekenstein2010,DeFelice2010,Linder2010,
Milgrom2014,Lovelock1971,Horava2009,Rodriguez2017,Horndeski1974,Deffayet2010,Deffayet2010}


Several popular modified gravity theories have been proposed to compete with the standard $\Lambda$CDM model, including Emergent Gravity \cite{Verlinde}, f(R) gravity \cite{Buchdahl1970} and scalar-tensor-vector gravity \cite{Moffat2006}. 
Therefore, it is very important to motivate some theoretical framework to differentiate the dynamical effects of the $\Lambda$CDM model and the modified gravity. Some recent studies proposed that using the turnaround radius (TAR) can be a clue to test the standard $\Lambda$CDM model and modified gravity models \cite{Bhattacharya2017,Lopes2018,Lopes2019,Pavlidou2014,Pavlidou2014a,Faraoni2015}. The TAR has been claimed to be a well-defined and unambiguous boundary of a structure (e.g. galaxy clusters) in simulations \cite{Pavlidou2014}. Different cosmological models and modified gravity models might have different general relations of the TAR. Therefore, determining the TAR of different structures precisely would be crucial to test and constrain different cosmological models \citep{Lopes2018}, DE, and disentangle between $\Lambda$CDM model, DE, and MG models \citep{Pavlidou2014,Pavlidou2014a,Faraoni2015,Bhattacharya2017,Lopes2018,Lopes2019}.

Contrarily to the previous claim, we already showed in \citep{DelPopolo2013a,DelPopolo2013b,Pace2014,Mehrabi2017,Pace2019,DelPopoloChan2020} that shear, and vorticity modifies the non-linear evolution of structures. In this paper, 
we will also show that dynamical friction further modifies the structure formation, and consequently modifies TAR, and that TAR generally depends on baryons physics \cite{DelPopoloChan2020}. By using an extended spherical collapse model, the TAR depends on the effects of shear and vorticity. Taking into account of the effects of shear and vorticity, the relation between TAR and total mass can differ by 30\% from that omit these effects, especially in galaxies \cite{DelPopoloChan2020}. In the present paper, we show that the effect of dynamical friction is also significant. 

TAR was calculated by \citep{Pavlidou2014,Pavlidou2014a} for the $\Lambda$CDM model smooth DE model, while \citep{Faraoni2015} obtained TAR in generic gravitational theories. 

In this paper, we extend the results of \citep{DelPopoloChan2020}, based on an extended spherical collapse model (ESCM) introduced, and adopted in   \citep{DelPopolo2013,DelPopolo2013a,Pace2014,Mehrabi2017,Pace2019}. 
The ESCM takes into account the effect of shear, vorticity and dynamical friction on the collapse, to show how the TAR is changed. 
Apart the typical parameters of the spherical collapse, shear, vorticity, and dynamical friction change the two-point correlation function \citep{DelPopolo1999}, the weak lensing peaks \citep{Pace2019}, and the mass function \citep{DelPopolo2013,DelPopolo2013a,Pace2014,Mehrabi2017}. Similarly, to \citep{DelPopoloChan2020}, the aim of the paper is to show how the parameters of the spherical collapse are changed, together with the $R_{\rm t}$-$M_{\rm t}$ relation for MG models, DE models, and to compare to $\Lambda$CDM model predictions.

{
The paper is organized as follows. 
Section~\ref{sect:model} describes the model used to derive the $R_{\rm t}$-$M_{\rm t}$ relation.
Section~\ref{sect:Results} is devoted to the discussion of our results.  
Section~\ref{sect:conclusions} is devoted to conclusions.
}

\section{The Model}\label{sect:model}

In the following, we will use an improved version \citep{Fillmore1984,Bertschinger1985,Hoffman1985,Ryden1987,Subramanian2000,
Ascasibar2004,Williams2004} of the spherical collapse model introduced by \cite{Gunn1972}. The model describes the evolution of perturbation from the linear to the non-linear phase, when them decouple from Hubble flow, reach a maximum radius, the TAR, collapse, and viriliaze forming a structure. As reported the initial model of \cite{Gunn1972} was extended to 
take account of angular momentum \citep{Ryden1987,Gurevich1988a,Gurevich1988b,White1992,Sikivie1997,Nusser2001,Hiotelis2002,  
LeDelliou2003,Ascasibar2004,Williams2004,Zukin2010}, of dynamical friction \citep{AntonuccioDelogu1994,Delpopolo2009}, shear  \citep{Hoffman1986,Hoffman1989,Zaroubi1993}, and the effects of the DE fluid perturbation   
\citep[see][]{Mota2004,Nunes2006,Abramo2007,Abramo2008,Abramo2009a,Abramo2009b,Creminelli2010,Basse2011,Batista2013}.
\cite{DelPopolo2013a,DelPopolo2013b} studied the effects of shear and rotation in smooth DE models, \cite{Pace2014b} in clustering DE cosmologies, and \cite{DelPopolo2013c} in Chaplygin cosmologies.

\subsection{The ESCM}

Here, we show how the evolution equations of $\delta$ in the non-linear regime can be obtained. 

The equations of evolution of $\delta$ in the non-linear regime were obtained by    \cite{Bernardeau1994,Padmanabhan1996,Ohta2003,Ohta2004,Abramo2007,Pace2010}. In order to obtain the equation, we used 
the Neo-Newtonian expressions for the relativistic Poisson equation, the Euler, and continuity equations \citep{Lima1997}
\begin{eqnarray}
  \frac{\partial\rho}{\partial t}+\nabla_{\vec{r}}\cdot(\rho\vec{v})+
  \frac{P}{c^2}\nabla_{\vec{r}}\cdot\vec{v} & = & 0 \label{eqn:cnpert}\;,\\
  \frac{\partial\vec{v}}{\partial
    t}+(\vec{v}\cdot\nabla_{\vec{r}})\vec{v}+
  \nabla_{\vec{r}}\Phi +{\bf \frac{c^2}{c^2 \rho+P} \nabla P} & = & 0\;, \label{eqn:enpert}\\
  \nabla^2\Phi-4\pi G\left(\rho+\frac{3P}{c^2}\right) & = & 0\;,\label{eqn:pnpert}
\end{eqnarray}
where the equation of state (EoS) is given by $P=w \rho c^2$, $\vec{r}$ indicates the physical coordinate, $\Phi$ the Newtonian gravitational potential, and $\vec v$ the velocity in three-space. 
Writing and combining the perturbation equation as in \citep{DelPopoloChan2020}, we obtain the non-linear evolution equation in a dust ($w=0$) universe 

\begin{equation}\label{eqn:nleq11}
\begin{split}
\ddot{\delta}+2H\dot{\delta}-\frac{4}{3}\frac{\dot{\delta}^2}{1+\delta}-
4\pi G\bar{\rho}\delta(1+\delta)-\\
(1+\delta)(\sigma^2-\omega^2) & = 0\;
\end{split}
\end{equation}
Eq. (\ref{eqn:nleq11}) is Eq.~41 of \cite{Ohta2003}, and a generalization of Eq.~7 of \cite{Abramo2007} to the case of a non-spherical configuration of a rotating fluid.

In Eq. (\ref{eqn:nleq11}), $H$ is the Hubble function, $\overline{\rho}=\rho-\delta \rho$ the background density,
$\sigma^2=\sigma_{ij}\sigma^{ij}$, and $\omega^2=\omega_{ij}\omega^{ij}$ are the shear, and rotation term, respectively. The shear term is related to a symmetric traceless tensor, dubbed shear tensor, while rotation term is related to an antisymmetric tensor.

In terms of the scale factor, $a$, the nonlinear equation driving the evolution of the overdensity contrast can be rewritten as:
\begin{equation}\label{eqn:wnldeq}
 \begin{split}
  \delta^{\prime\prime}+\left(\frac{3}{a}+\frac{E^\prime}{E} \right)
\delta^\prime-\frac{4}{3}\frac{\delta^{\prime 2}}{1+\delta}-
  \frac{3}{2}\frac{\Omega_{\mathrm{m},0}}{a^5 E^2(a)}\delta(1+\delta)-&\\
  \frac{1}{a^2H^2(a)}(1+\delta)(\sigma^2-\omega^2)&=0\;,
 \end{split}
\end{equation}
where $\Omega_{m,0}$ is DM density parameter at $t=0$ ($a=1$), and $E(a)$ is given in Eq. 11 of \citep{DelPopoloChan2020}.

{
Since $\delta=\frac{2GM_{\rm m}}{\Omega_{m,0} H^2_0}(a/R)^3-1$, where $R$ is the effective perturbation radius, inserting into 
Eq.~(\ref{eqn:nleq11}), it is easy to check that the evolution equation for $\delta$ reduces to the spherical collapse model (SCM) \citep{Fosalba1998a,Engineer2000,Ohta2003}

\begin{eqnarray} \label{eqn:wnldeq0}
 \ddot{R} &=& -\frac{GM_{\rm m}}{R^2} - \frac{GM_{\rm de}}{R^2}(1+3w_{\rm de})-\frac{\sigma^2-\omega^2}{3}R= \nonumber\\
& & 
-\frac{GM_{\rm m}}{R^2} - \frac{4 \pi G \bar{\rho_{\rm de}} R}{3}(1+3w_{\rm de})-\frac{\sigma^2-\omega^2}{3}R\,,
\end{eqnarray}
where $M_{\rm de}$ is the mass of the dark-energy component enclosed in the volume, $M_{\rm m}= \frac{4 \pi R^3}{3} (\bar{\rho}+\delta \rho)$, $w_{\rm de}$, and $\bar{\rho}_{\rm de}$, are the DE equation-of-state parameter and its background density, respectively \citep{Fosalba1998a,Engineer2000,Ohta2003,Pace2019}.
}

In the case $w=-1$, namely the cosmological constant, $\Lambda$, case, Eq. (\ref{eqn:wnldeq0}), for $w=-1$, can be written as
\begin{equation} \label{eqn:spher0}
 \ddot{R}=-\frac{GM_{\rm m}}{R^2} -\frac{\sigma^2-\omega^2}{3}R+ \frac{\Lambda}{3} R\
\end{equation}
The previous equation is clearly similar to the usual expression for the SCM with cosmological constant, and angular momentum \citep[e.g.][]{Peebles1993,Nusser2001,Zukin2010}:
\begin{equation}
\frac{d^2 R}{d t^2}= -\frac{GM}{R^2} +\frac{L^2}{M^2 R^3}+\frac{\Lambda}{3}R= -\frac{GM}{R^2} + \frac{4}{25} \Omega^2 R+ \frac{\Lambda}{3}R,
\label{eqn:spher}
\end{equation}
The last right term in Eq. \ref{eqn:spher} is obtained recalling that $L=I \Omega$, and the 
momentum of inertia of a sphere, $I=2/5 M R^2$.

{
Angular momentum is related to vorticity by $\Omega=\omega/2$ (see also \cite{Chernin1993}), in the case of a uniform rotation with angular velocity $\Omega=\Omega_{\rm z} {\bf e}_{\rm z}$.
As in \citep{DelPopolo2013a,DelPopolo2013b,DelPopoloChan2020}, we define the dimensionless, but mass dependent, quantity $\alpha$ as the ratio between the rotational and the gravitational term in Eq. (\ref{eqn:spher}):
\begin{equation}
\alpha(M)=\frac{L^2}{M^3RG}
\end{equation}

In order to solve Eq. (\ref{eqn:nleq11}), the relation between the term $\sigma^2-\omega^2$, and the density contrast, $\delta$ is needed. This connection can be obtained recalling the relation between angular momentum and shear, and recalling that Eq. (\ref{eqn:spher}) from which $\alpha$ was obtained is equivalent to Eq. (\ref{eqn:wnldeq0}) which is also equivalent to Eq. (\ref{eqn:nleq11}).

Calculating the same ratio between the gravitational and the extra term appearing in Eq. (\ref{eqn:nleq11}) we obtain
\begin{equation} \label{eq:nonlinf}
\frac{\sigma^2-\omega^2}{H^2_0}=-\frac{3}{2}\frac{\alpha \Omega_{\rm m,0}}{a^3}\delta.
\end{equation}

This reasonable assumption (see \citep{DelPopolo2013b}) was also used in \citep{DelPopolo2013a,DelPopolo2013b,Pace2014,Mehrabi2017}. 
%
%
}

The nonlinear equation to solve is obtained substituting Eq. (\ref{eq:nonlinf}) into Eq. (\ref{eqn:nleq11})
\begin{equation} \label{eq:centr}
\begin{split} 
\ddot{\delta}+2H\dot{\delta}-\frac{4}{3}\frac{\dot{\delta}^2}{1+\delta}-
4\pi G\bar{\rho}\delta(1+\delta)-\\
-\frac{3}{2} H_0^2 (1+\delta) \frac{\alpha \Omega_{\rm m,0}}{a^3}\delta  & = 0\;
\end{split}
\end{equation}


Solving Eq. (\ref{eq:centr}) following the method described in \citep{Pace2010}, or solving Eq. (\ref{eqn:wnldeq0}),
the threshold of collapse, and the turnaround, can be obtained.

At this point, we want to take into account also dynamical friction in our analysis. Then we notice that Eq. (\ref{eqn:wnldeq0}), can be written in a more general form taking into account dynamical friction
 \citep{Kashlinsky1986,Kashlinsky1987,Lahav1991,Bartlett1993,
 AntonuccioDelogu1994,Peebles1993,DelPopolo1998,DelPopolo1998a,DelPopolo2006b,Delpopolo2009,DelPopolo2019}
\begin{equation}\label{eqn:wnldeqqq}
\begin{split}
 \ddot{R} = -\frac{GM}{R^2} + \frac{L^2(R)}{M^{2}R^3} + \frac{\Lambda}{3}R -
 \eta\frac{{\rm d}R}{{\rm d}t}=\\
 -\frac{GM_{\rm m}}{R^2} - \frac{GM_{\rm de}}{R^2}(1+3w_{\rm de})-\frac{\sigma^2-\omega^2}{3}R -\eta \frac{dR}{dt}.
 \end{split}
\end{equation}
being $\eta$ the dynamical friction coefficient. 
Eq. (\ref{eqn:wnldeqqq}) can be obtained via Liouville's theorem \citep{DelPopolo1999}, and the dynamical friction force per unit mass, $\eta\frac{{\rm d}R}{{\rm d}t}$, is given in \citep{Delpopolo2009}(Appendix D, Eq. D5), and \citep{DelPopolo2006b}, Eq. 5).

A similar equation (excluding the dynamical friction term) was obtained by several authors 
\citep[e.g.,][]{Fosalba1998a,Engineer2000,DelPopolo2013b}) and generalized to smooth DE models in 
\cite{Pace2019}.

Eq. (\ref{eqn:wnldeqqq}), and Eq. (\ref{eqn:wnldeq0}) differs for the presence of the dynamical friction term. 
Dynamical friction similarly to rotation, and cosmological constant delays the collapse of a structure (perturbation).
\citep{AntonuccioDelogu1994,Delpopolo2009,DelPopolo2006b,DelPopolo2017,DelPopolo2019}. The magnitude of the effect of cosmological constant, rotation, and dynamical friction, are of the same order with differences of a few percent (see Fig. 1 of \citep{DelPopolo2017}, and Fig. 11 of \citep{Delpopolo2009}).

Notice that, by means of the relation $\delta=\frac{2GM_{\rm m}}{\Omega_{m,0} H^2_0}(a/R)^3-1$, Eq. (\ref{eqn:wnldeqqq}) can be written in terms of $\delta$, similarly to Eq. (\ref{eqn:nleq11}).

%
%

Our SCM model depends not only from shear, vorticity, as in \citep{DelPopoloChan2020}, but also on dynamical friction.
Since shear, rotation, and dynamical friction depends from the mass, the SCM results depend from the baryon physics, differently from what claimed by \citep{Pavlidou2014,Lopes2018,Bhattacharya2017}.

\section{Results}\label{sect:Results}

\begin{figure*}[!ht]
 \centering
 \includegraphics[width=18cm,height=6cm,angle=0]{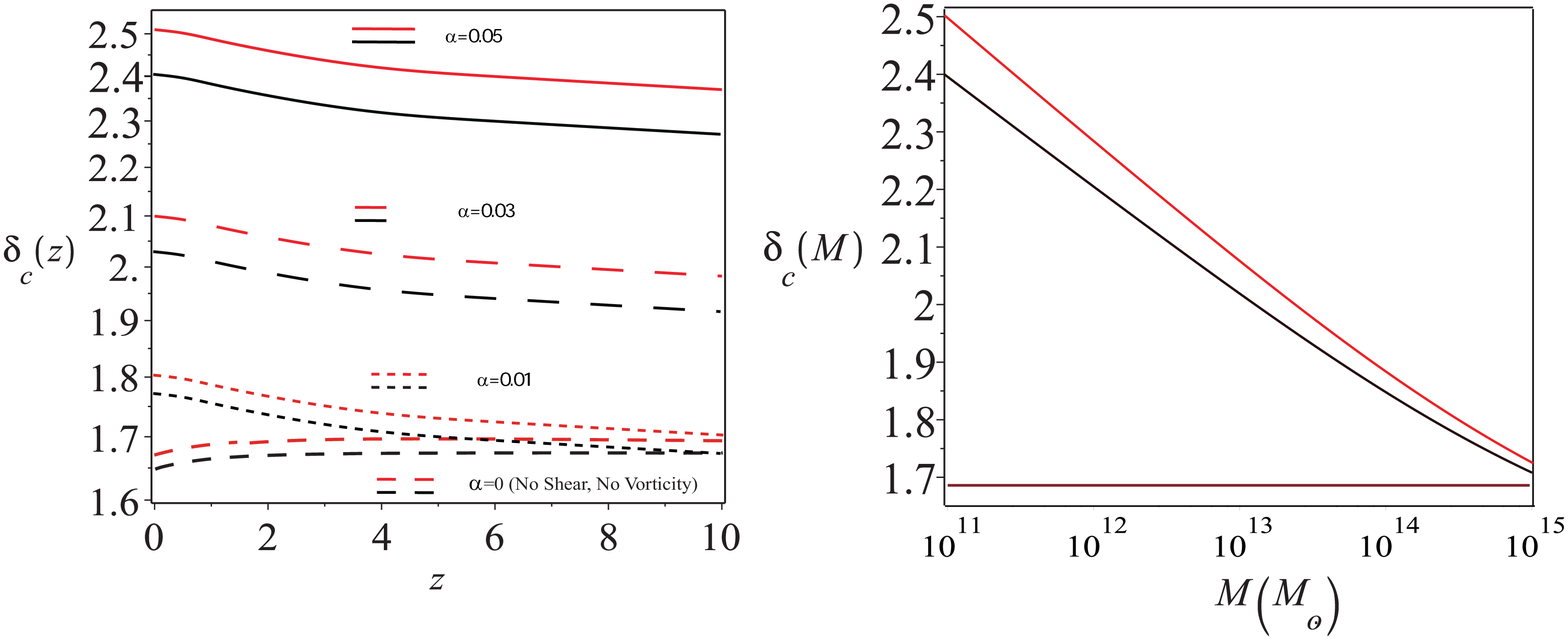} 
 \caption[justified]{The threshold of collapse $\delta_c$, as function of redshift, and mass.
 {In the left panel, $\delta_c$ vs redshift is plotted. The prediction for $\delta_c(z)$ for the $\Lambda$CDM model are represented by the red lines. The values of $\alpha$ varies from 0.05 (solid line) (galactic mass scale), to 0.03 (long dashed line), to 0.01 (dotted line) (cluster mass scale), and to 0 (dashed line). The right panel plots $\delta_c$ vs mass. The solid red line represents the result of the ESCM model, the red dashed line that of {the elliptictal collapse model of} \citep{Sheth2001}, and the brown line the $\Lambda$CDM expectation when no shear and rotation are taken into account.}}
 \label{fig:comparison}
\end{figure*}

The effect of shear \citep{Hoffman1986,Hoffman1989, Zaroubi1993}, and rotation \citep{DelPopolo1998,DelPopolo1999,DelPopolo2000,DelPopolo2001,DelPopolo2002,DelPopolo2013b,Pace2019} on the collapse are manifold. 

One general feature is that of slowing down the collapse \citep{Peebles1990,Audit1997,DelPopolo2001,DelPopolo2002}. Mass function 
\citep{DelPopolo1999,DelPopolo2000,DelPopolo2013b,Pace2014,Mehrabi2017,DelPopolo2017,Pace2019}, two-point correlation function, \citep{DelPopolo2005}, scaling relations like the mass-temperature, and luminosity-temperature relation \citep{DelPopolo2005} \citep{DelPopolo2002a,DelPopolo2019}, are modified. This is connected to the change of the typical parameters of the SCM.

\subsection{Threshold of collapse with shear, rotation, and dynamical friction}

\begin{figure*}[!ht]
 \centering
\includegraphics[width=10cm,angle=0]{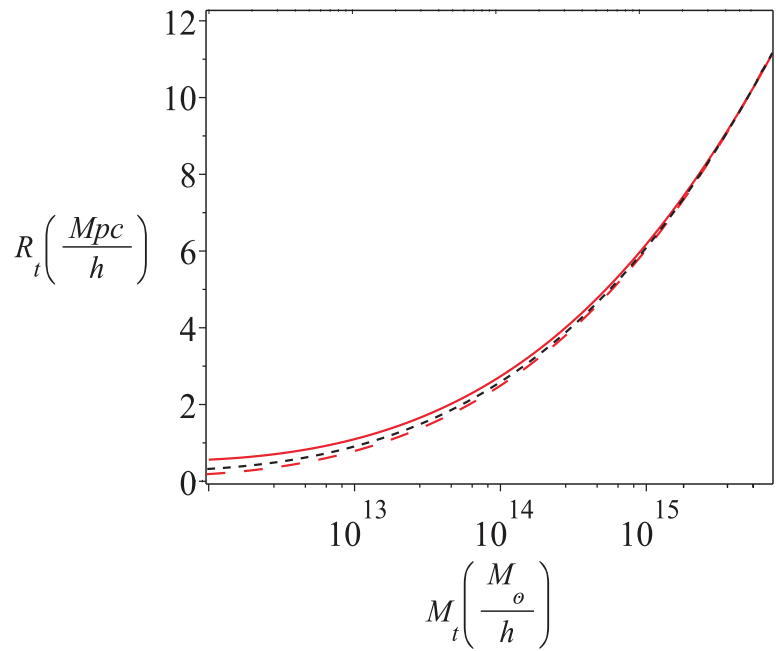}  
 \caption[justified]{Turnaround radius $R_{\rm t}$ vs mass, $M_{\rm t}$. The solid red line represents the TAR predicted by the standard SCM, and the dashed red line ESCM result for the $\Lambda$CDM model. The black dotted line
the result of the ESCM for the AS model. 
}
 \label{fig:comparison}
\end{figure*}

Fig. 1 shows how shear, rotation, and dynamical friction (shortly SRD) change the collapse threshold,  $\delta_c$. 
We show the dependence from redshift in the left panel, and mass in the right panel.
The red line represents the predictions of the ESCM for  $\delta_c(z)$ for the $\Lambda$CDM model, while the black line  
the ESCM in the case of one DE model, the Albrecht-Skordis \citep{Albrecht2000} (AS) (black line).

In the left panel, from top to bottom, the value of $\alpha$ varies from 0.05 (solid line) corresponding to a mass $\simeq 10^{11} M_{\odot}$, to 0.03 (long dashed line), corresponding to a mass $\simeq 10^{13} M_{\odot}$, to 0.01 (dotted line), corresponding to a mass $\simeq 10^{15} M_{\odot}$, and to 0 (dashed line). The value of $\delta_c(z)$ for $\alpha=0.05$, is $\simeq 30\%$ larger than in the $\alpha=0$ case.

In the case, of no shear, rotation ($\alpha=0$) and dynamical friction, $\delta_c$, has a weak dependence from redshift in the $z$ range $[0,2]$, and then assumes the value predicted by the Einstein de Sitter model. 

In other terms, SRD gives rise to a non-flat threshold, $\delta_c$, which is monotonically decreasing with redshift.
Moreover, the larger is $\alpha$, the larger is the difference between the values of $\delta_c(z)$, as shown by the different curves.

Of the DE quintessence models in literature, we plotted the AS model because the other models considered in previous papers \citep{Pace2010,DelPopolo2013a} (INV1 ($w_0=-0.4$), INV2 ($w_0=-0.79$), 2EXP ($w_0=-1$), CNR ($w_0=-1$), CPL ($w_0=-1$), SUGRA ($w_0=-0.82$) (see \citep{Pace2010,DelPopolo2013b})\footnote{$w_0$ is the value of $w$ nowadays})
are contained in the envelope between the region included in the $\Lambda$CDM and the AS model (see Fig. 4 of \citep{Pace2010}).

The right panel of Fig. 1 plots $\delta_c(M)$ versus the mass. In absence of SRD the value of $\delta_c$ is constant (brown line).
in presence of SRD, $\delta_c$ becomes mass dependent, and monotonically decreases with mass. This means that in order a less massive perturbations (e.g., galaxies) form structures must cross a higher threshold than more massive ones. This behavior, is related to the 
anticorrelation of the angular momentum acquired by the proto-structure and its height\footnote{The peak height is defined as $\nu=\delta_c/\sigma(M)$, where $\sigma(M)$ is the mass variance. We have that the specific angular momentum, $j$ is given by $j \propto \nu^{-3/2}$ \citep{Hoffman1986,Delpopolo2009,Polisensky2015}}. 
Since low peaks acquire larger angular momentum than high peaks, they need a higher density contrast to collapse and form structures \citep{DelPopolo1998,DelPopolo2001,DelPopolo2002,Delpopolo2009,Ryden1988,Peebles1990,Audit1997}.

As shown in the left and right panel of Fig. 1, the effect of dynamical friction is that of increasing the values of $\delta_c(z)$, and $\delta_c(M)$ with respect to the case it is not present as shown in \citep{DelPopoloChan2020}.
 

\begin{figure*}[!ht]
 \centering
  \includegraphics[width=12cm,angle=0]{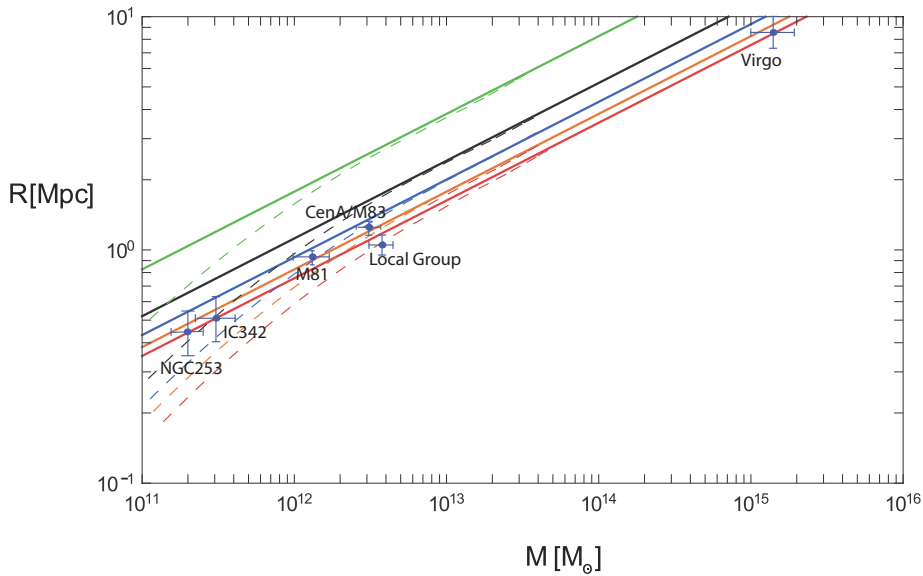}
 \caption[justified]{Stable structures mass-radius relation for different $w$. 
 The solid lines from top to bottom represent $w=-0.5$ (solid green line), -1 (black solid line), -1.5 (blue solid line) ,-2 (pink solid line), -2.5 (red solid line). The results of the ESCM are represented by the dashed lines. The dots with error bars, are data as in \citep{Pavlidou2014a}.}
 \label{fig:comparison}
\end{figure*}

\subsection{Comparison of TAR, in the $\Lambda$CDM, ESCM, and DE models}

Shear, rotation, and dynamical friction modify the TAR. In order to show this, we may compare the predictions of the $\Lambda$CDM, that of the ESCM, and DE models. \citep{Pavlidou2014} and \citep{Pavlidou2014a} calculated the maximum TAR, MTAR, that is, the radius of the surface where $\ddot R =0$. \citep{Pavlidou2014a} found  
\begin{equation} \label{eq:pavl}
R_{\rm ta}= \left [\frac{-3M}{4 \pi \rho_{\rm de}(1+3w)} \right]^{1/3}
\end{equation}
which in the case of the $\Lambda$CDM model ($w=-1$) reduces to
\begin{equation} \label{eq:turn1}
R_{\rm ta}= \left [\frac{3GM}{\Lambda} \right]^{1/3}
\end{equation}
\citep{Pavlidou2014}.

In their estimation, they assumed that shear and rotation were not present. Their expression can be generalized to the case shear, and rotation are non zero. This can be done using Eq. (\ref{eqn:wnldeq0}), obtaining
\begin{equation} \label{eq:turnn}
R_{\rm ta}= \left [\frac{-3M}{4 \pi \rho_{\rm de}(1+3w)+(\sigma^2-\omega^2)} \right]^{1/3}
\end{equation}

{In the paper, we will get and plot the TAR, not MTAR, since we compare with \citep{Lopes2018}, which calculated the TAR.}

Fig. 2, shows the TAR predicted by the $\Lambda$CDM model (solid red line), that of the ESCM model (taking into account shear rotation, and dynamical friction) (red dashed line) for the $\Lambda$CDM model, and the AS model (black dashed line). 
When shear, rotation and dynamical friction are taken into account the collapse is slowed down, and the the TAR is smaller.
The difference between the $\Lambda$CDM model, and the ESCM preditions increases going towards smaller masses. This is mainly related to the larger rotation of smaller objects, and reach a maximum difference of $\simeq 60\%$. 
%
%
The black dashed line, as already reported, is the AS model, which has a slightly larger TAR with 

\subsection{Constraints on DE EoS parameter}

\begin{table}
\caption{The allowed ranges of $w$ based on the ESCM model.}
 \label{table1}
 \begin{tabular}{@{}ll}
  \hline
  Stable structure  & range of $w$  \\
  \hline
  M81     &$w \ge -1.5$\\
  IC342   &$w \ge -1$\\
  NGC253  &$w \ge -1$\\
  CenA/M83    &$w \ge -1.5$\\
  Local Group &$w \ge -2$\\
  Virgo       &$w \ge -2$\\
  \hline
 \end{tabular}
\end{table}

%
%

\begin{figure*}[!ht]
 \centering
\includegraphics[width=12cm,angle=0]{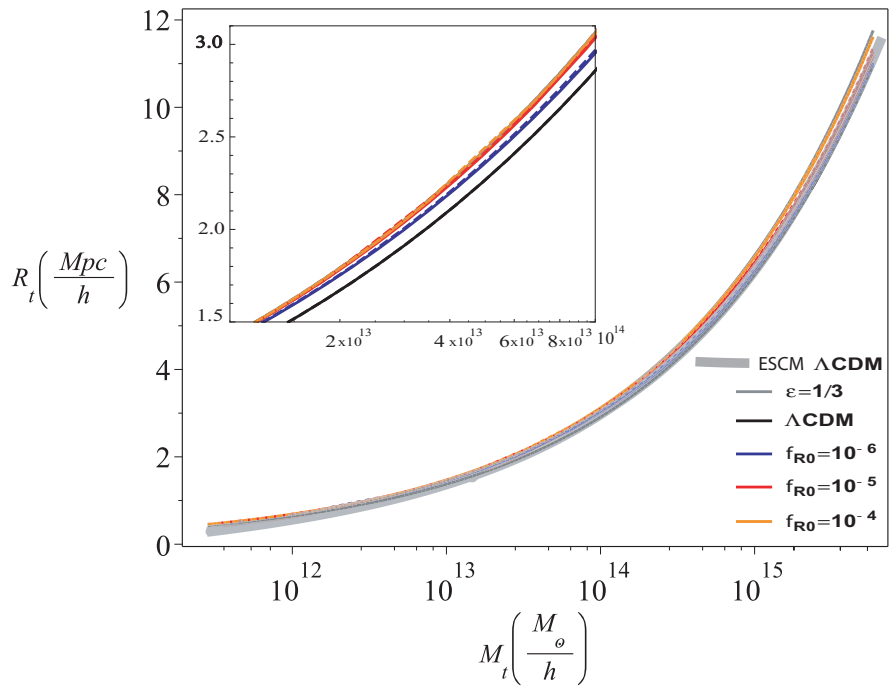} 
 \caption[justified]{Turnaround radius, $R_{\rm t}$ vs mass, $M_{\rm t}$. The plot represents the result obtained by \citep{Lopes2018} for $R_{\rm ta}$ vs mass for the $\Lambda$CDM model, the model with $\epsilon=1/3$, and $f(R)$ with $f_{\rm R0}=10^{-6}$, $10^{-5}$, and $10^{-4}$. The gray band represents our prediction for the $\Lambda$CDM model, obtained with the ESCM, with the 68\% confidence level region.}
 \label{fig:comparison}
\end{figure*}

\begin{figure*}[!ht]
 \centering
 \includegraphics[width=18cm,angle=0]{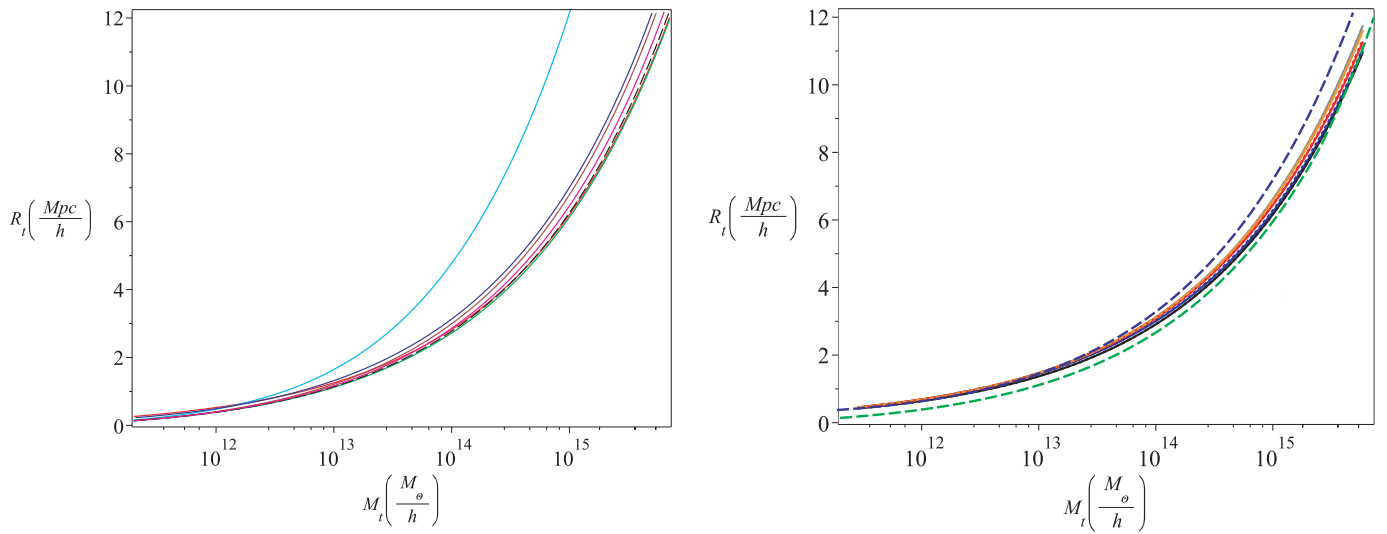}
 \caption[justified]{The turnaround radius, $R_{\rm t}$ vs mass, $M_{\rm t}$ obtained with the ESCM for the DE models. Left panel: from top to bottom, the cyan, blue, brown, magenta, black, red, and green lines represent the INV1, INV2, SUGRA, $w_{09}$\footnote{Namely the model having $w=-0.9$}, AS, $\Lambda$CDM without shear, and rotation, and $\Lambda$CDM with shear, and rotation, respectively. Right panel: the solid lines are the $\Lambda$CDM, and $f(R)$ models of \citep{Lopes2018}, 
while the blue dashed line, and the green dashed lines, are the INV2 model, and the $\Lambda$CDM with shear, rotation, and dynamical friction, plotted in the left panel.}
 \label{fig:comparison}
\end{figure*}

Knowing the relation between TAR, the structure mass, and the EoS parameter, it is possible to obtain some constraints on the DE EoS ($w$). This was tried by \citep{Pavlidou2014a} by means of Eq. (\ref{eq:pavl}). In general, a great number of DE models are described through the equation of state parameter, $w(a)$. The last depends on its present value $w_0$, on that at matter-radiation equality epoch, and some other parameters at the same epoch (see Eq. 23 of \citep{Pace2010}). If one wants to constrain the EoS parameter, $w$ evolution, high redshift structures are needed. If by converse, one wants to constrain the values of $w_0$ 
small $z$ cosmic structures may be used. \citep{Pavlidou2014a} compared the predicted TAR at different $w$ in the plane $M_t$-$R_t$, finding constraints on $w_0$. To this aim data from several structures (e.g., Milky Way (MW), M81/M82 group, Local Group, Virgo cluster, Fornax-Eridanus group) were used. It is of fundamental importance to note, that the constrain depends from the model used. 
As shown by \citep{Peirani2008}, the mass of the structure, and its TAR changes when to the standard SCM (considering only the gravitational potential) is also added the cosmological constant (Eq. 1 of \citep{Peirani2008}). This was also pointed out comparing the values predicted by several Karachentsev's paper (e.g., \citep{Karachentsev2002,Karachentsev2005}, for M81, the local group, and neighboring groups) with those of \citep{Peirani2008}. 

%
%

In Fig. 3, the solid lines obtained from the equation of TAR (Eq. \ref{eq:pavl}), correspond to $w=-2.5$, -2, -1.5, -1, -0.5, from bottom to top when shear, rotation, and dynamical friction are absent. The dashed lines are the corrections obtained when shear, rotation, and dynamical friction are taken into account. The range of $w$ for which no stable structures exist is given by the parameter space above each line. \citep{Pavlidou2014a} discussed some constraints to $w$, based on the highest mass objects. As shown by the dashed lines, one has to expect that at masses smaller than $10^{13} M_{\odot}$ the TAR is modified by the presence of shear, rotation, and dynamical friction. As a consequence, structures at smaller masses can give different constraints to $w$. At the same time, following \citep{Peirani2008}, taking also the effect of dynamical friction, we see that the constraints on the cosmic structure studied, and plotted in Fig. 3, is noteworthy different from that of \citep{DelPopoloChan2020}. The values of TAR and mass for each of the objects in Fig. 3 were obtained using the SCM with dynamical friction, and using a method described by \citep{Peirani2006,Peirani2008} (see Appendix).
In Table 1, we report the constraints we obtained.

\subsection{Comparison with TAR in $f(R)$ theories}

In this section, we compare the evolution of the TAR in the ESCM, and the $f(R)$ theories. The evolution of TAR in  
General Relativity (GR), and in the $f(R)$ theories was investigated by \citep{Lopes2018}. Modified Gravity (MG) effects were introduced in the equation of the evolution of overdensity (their Eq. 3.3) by means of the parameter $\epsilon(a,k)$, where $k$ is the (angular) wavenumber. GR is recovered when $\epsilon(a,k)$ is zero, and our Eq. (\ref{eqn:wnldeq}), is recovered in the case $\epsilon(a,k)$ is zero, and shear, and rotation are put equal to zero. In other terms, our Eq. (\ref{eqn:wnldeq}) is a generalization of Eq. 3.3 of \citep{Lopes2018}, for the case $\epsilon(a,k)$ is zero.
Consequantly, their Figs. 1, 3, similarly to the left panel of our Fig. 1, and to the $\Lambda$CDM, and DE models without shear, and rotation, as shown in \citep{Pace2010}, shows  a monotonic increase of $\delta_c(z)$. Their Fig. 4 shows an almost flat behavior of $\delta_c(M)$, with variations from constancy of the order of 1\%.  
The behavior of $\delta_c(z)$, and $\delta_c(M)$ in \citep{Lopes2018} disagrees with the prediction of several papers (e.g.,  \citep{Sheth2001,DelPopolo2013a,DelPopolo2013b,Pace2014,Mehrabi2017,DelPopolo2017}). The previous papers showed that in order to have a mass function reproducing simulations, the threshold must be a monotonic decreasing function of mass. Since in \citep{Lopes2018}  
$\delta_c(M)$ is practically constant, and $\delta_c(z)$ is a monotonic increasing function of redshift, this implies that \citep{Lopes2018} results cannot reproduce the mass function obtained in simulations, and observations. This is due to the 
the fact \citep{Lopes2018} discarded the effect of shear, and rotation, and in general of aspericities.
A more detailed discussion of this aspect can be found in \citep{DelPopoloChan2020}.

~\\

Going back to our main goal, namely the use of the TAR to disentangle between GR and MG, and GR and DE models, discussed by several authors (e.g., \citep{Lopes2018}), we will compare \citep{Lopes2018} prediction on the $M_t$-$R_t$ relation with that of our ESCM. 
In Fig. 4, the result obtained by \citep{Lopes2018} for $R_{\rm t}$ vs mass for the $\Lambda$CDM model, the model with $\epsilon=1/3$, and $f(R)$ with $f_{\rm R0}=10^{-6}$, $10^{-5}$, and $10^{-4}$, are plotted. 
Our prediction for the $\Lambda$CDM model, obtained with the ESCM, with the 68\% confidence level region, are represented by the grey band. The 68\% confidence level was obtained, as in \citep{DelPopoloChan2020}, by means of a Monte Carlo simulation.
The result of the plot is slightly different from Fig. 4 of \citep{DelPopoloChan2020}. Because of the presence of dynamical friction, which further contribute to slow down the collapse, the TAR has smaller values.  
Consequently, the TAR in the ESCM model does not completely overlap with that of \citep{Lopes2018}, as happened in \citep{DelPopoloChan2020}. This means that the study of the top values of the $M_t$-$R_t$ relation could disentangle the GR predictions from that of the $f(R)$ theories. Choosing peculiar values of the TAR would be possible to  
disentangle between GR, and $f(R)$ theories. 
%
%

The predictions of some of the quintessence DE models previously cited 
are compared in Fig. 5 to the same $f(R)$ models of \citep{Lopes2018} plotted in Fig. 4. 
All the curves are obtained by means of the ESCM applied to the DE models of the TAR.
From top to bottom, the cyan, blue, brown, magenta, black, red, and green lines represent the INV1, INV2, SUGRA, $w_{09}$\footnote{Namely the model having $w=-0.9$}, AS, $\Lambda$CDM without shear, and rotation, and $\Lambda$CDM with shear, and rotation, respectively.

The comparison of the \citep{Lopes2018} predictions for TAR  with that of the DE models are plotted in the right panel. In this plot, we show only INV2 (blue dashed line), and the $\Lambda$CDM with shear, rotation, and dynamical friction (green dashed line). With exception of INV1, the two quoted curves contain all other DE models. 

The 
other lines are the $\Lambda$CDM, and $f(R)$ models of \citep{Lopes2018}. Differently from Fig. 4, the DE models contain the \citep{Lopes2018} prediction for TAR. As a consequence the TAR cannot be used to disentangle DE, and $f(R)$ theories predictions.

%
%

\section{Conclusions}\label{sect:conclusions}

In this paper, we discussed how shear, rotation, and dynamical friction change the TAR, and 
some of the parameters of the SCM. The results were obtained using an ESCM taking into account the effects of shear, vorticity, and dynamical friction, to determine the $R_{\rm t}-M_{\rm t}$, in $\Lambda$CDM, and in DE scenarios. We extended numerically the formula for maximum TAR obtained in \citep{DelPopoloChan2020}, 
to take into account dynamical friction. The value of TAR is reduced by shear, rotation, and dynamical friction, especially at galactic scales. 
Using the $R_{\rm t}-M_{\rm t}$ relationship, and data from stable structures, one can obtain constraints to $w$. 
Its values are smaller for structures with masses approximately smaller that $10^{13} M_{\odot}$. In this paper, we recalculated the mass, and TAR of M81/M82 group, Local Group, Virgo cluster, NGC253, IC342, CenA/M83 group following \citep{Peirani2006,Peirani2008}. 

A comparison of the $R_{\rm t}-M_{\rm t}$ relationship obtained for $\Lambda$CDM, and DE scenarios with \citep{Lopes2018} prediction of the $f(R)$ theories, shows that $R_{\rm t}-M_{\rm t}$ relationship in the $f(R)$ models are practically identical to that of 
of DE scenarios. This implies that the $R_{\rm t}-M_{\rm t}$ relationship is not a good probe to disentangle between GR, and DE models predictions. The situation is different in the case of the $\Lambda$CDM model. In this case, the 68\% confidence level region
does not overlap with that of the $f(R)$ models. The higher values of the TAR could be used to disentangle between $f(R)$ theories, and GR.
%
%


\appendix

\section{Mass and TAR of structures}

The most general equation taking account of shear, rotation, and dynamical friction is Eq. (\ref{eqn:wnldeqqq}). 
We will rewrite it in adimensional form. Assuming that $J=k R^{\alpha}$, with $\alpha=1$, in agreement with \citep{Bullock2001}\footnote{In that paper $\alpha=1.1 \pm 0.3$}, and $k$ constant. In terms of the variables $y=R/R_0$, $t=x/H_0$, 
Eq. (\ref{eqn:wnldeqqq}) can be written as
\begin{equation}
\label{eq:princ}
\frac{d^2y}{dx^2}=-\frac{A}{2y^2}+\Omega_{\Lambda} y+\frac{K_j}{y}-\frac{\eta}{H_0}\frac{dy}{dx},
\end{equation}
where $K_j=k\frac{1}{(H_0R_0)^2}$\footnote{$K_j=0.78$, and $\frac{\eta}{H_0}=0.5$}, $A=\frac{2GM}{H_0^2 R_0^3}$, and
\begin{equation}
\label{eq:hubb1}
H=H_0 \sqrt{\left(\frac{a_0}{a}\right)^3 \Omega_m+\Omega_{\Lambda}}.
\end{equation}
Eq.\eqref{eq:princ} has a first integral, given by
\begin{align} 
\label{eq:princ1}
&u^2=\left(\frac{dy}{dx}\right)^2\nonumber\\
&=\frac{A}{y}+\Omega_{\Lambda} y^2+2K_j \log{y}-2\frac{\eta}{H_0} \int \left(\frac{dy}{dx}\right)^2 dx+K
\end{align}
where $K=\frac{2E}{(H_0R_0)^2}$, and $E$ is the energy per unit mass of a shell.

The previous equation can solved as described in \citep{Peirani2006,Peirani2008}. 
%
%

\begin{figure}[t]  
 \centering
\includegraphics[scale=0.45]{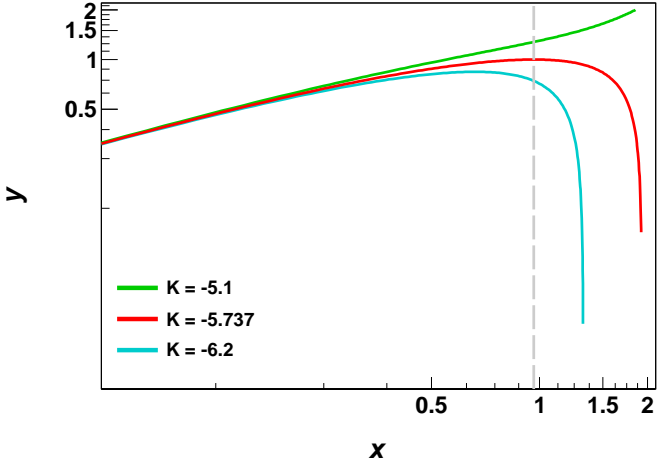}
\caption[justified]{
	Evolution of shell radius for different values of $K$. The red, cyan, and green lines correspond to $K=-5.737$, $K=-6.2$, and $K=-5.1$, respectively.
}
\label{fig:RadiusEvoluion_diff_K}
\end{figure}

The mass, and turn-around radius of some groups of galaxies, is obtained finding a relation between the velocity, and radius, $v-R$. The last will be fitted to the data. The $v-R$ relation is obtained as follows. Let's consider Fig.\ref{fig:RadiusEvoluion_diff_K}. This is a solution of Eq. (\ref{eqn:wnldeqqq}) for different values of $K$. The vertical line corresponds to $x=0.964$. Its intersection with the curves, solution of Eq. (\ref{eqn:wnldeqqq}) (cyan, red, green) gives, for each one a value $y(x)=y(0.964)$. The solution of Eq. (\ref{eqn:wnldeqqq}), also gives the velocity, allowing us to find $u(x)=u(0.964)$. We will get a couple of value $(y,u)$ for each intersection of the vertical line with the curves. This allows us to find a series of points that can be fitted with a relation of the form $u=-b/y^n+by$ obtaining $u=-1.3436/y^{0.9107}+1.3436y$.
This last relation can be written in physical units as follows.  
\begin{equation} 
\label{eq:recov}
v(R)=-b H_0 R_0 \left(\frac{R_0}{R}\right)^n+b H_0 R
\end{equation}. 
Substituting in this equation, $R_0=(\frac{2GM}{H_0^2})^\frac{1}{3}$, we get
\begin{equation}
v(R)=-b\frac{H_0}{R^n} \left(\frac{2GM}{A H_0^2}\right)^\frac{n+1}{3}+bH_0R
\end{equation}
or
\begin{equation}
\label{eq:eta}
v(R)=-\frac{-0.66385 H_0}{R^n} \left(\frac{GM}{H_0^2}\right)^\frac{n+1}{3}+1.3436 H_0 R
\end{equation}
where $n=0.9107$. Eq. (\ref{eq:eta}) satisfy the condition $v(R_0)=0$.
Fitting the equation to the data of \citep{Peirani2006} (Fig. 2), and \citep{Peirani2008} (Fig. 2) one obtains the value of the Hubble parameter , $H_0$, and TAR. Solving  Eq. (\ref{eqn:wnldeqqq}) one can obtain the value of $A$, and solving the equation $A=\frac{2GM}{H_0^2 R_0^3}$, being $H_0$, and $R_0$, one gets the mass, $M$.

%
%

\begin{thebibliography}{113}%
\makeatletter
\providecommand \@ifxundefined [1]{%
 \@ifx{#1\undefined}
}%
\providecommand \@ifnum [1]{%
 \ifnum #1\expandafter \@firstoftwo
 \else \expandafter \@secondoftwo
 \fi
}%
\providecommand \@ifx [1]{%
 \ifx #1\expandafter \@firstoftwo
 \else \expandafter \@secondoftwo
 \fi
}%
\providecommand \natexlab [1]{#1}%
\providecommand \enquote  [1]{``#1''}%
\providecommand \bibnamefont  [1]{#1}%
\providecommand \bibfnamefont [1]{#1}%
\providecommand \citenamefont [1]{#1}%
\providecommand \href@noop [0]{\@secondoftwo}%
\providecommand \href [0]{\begingroup \@sanitize@url \@href}%
\providecommand \@href[1]{\@@startlink{#1}\@@href}%
\providecommand \@@href[1]{\endgroup#1\@@endlink}%
\providecommand \@sanitize@url [0]{\catcode `\\12\catcode `\$12\catcode
  `\&12\catcode `\#12\catcode `\^12\catcode `\_12\catcode `\%12\relax}%
\providecommand \@@startlink[1]{}%
\providecommand \@@endlink[0]{}%
\providecommand \url  [0]{\begingroup\@sanitize@url \@url }%
\providecommand \@url [1]{\endgroup\@href {#1}{\urlprefix }}%
\providecommand \urlprefix  [0]{URL }%
\providecommand \Eprint [0]{\href }%
\providecommand \doibase [0]{http://dx.doi.org/}%
\providecommand \selectlanguage [0]{\@gobble}%
\providecommand \bibinfo  [0]{\@secondoftwo}%
\providecommand \bibfield  [0]{\@secondoftwo}%
\providecommand \translation [1]{[#1]}%
\providecommand \BibitemOpen [0]{}%
\providecommand \bibitemStop [0]{}%
\providecommand \bibitemNoStop [0]{.\EOS\space}%
\providecommand \EOS [0]{\spacefactor3000\relax}%
\providecommand \BibitemShut  [1]{\csname bibitem#1\endcsname}%
\let\auto@bib@innerbib\@empty
\bibitem [{\citenamefont {{Sanders}}(2010)}]{Sanders}%
  \BibitemOpen
  \bibfield  {author} {\bibinfo {author} {\bibfnamefont {R.~H.}\ \bibnamefont
  {{Sanders}}},\ }\href@noop {} {\emph {\bibinfo {title} {The Dark Matter
  Problem: A Historical Perspective}}}\ (\bibinfo  {publisher} {Cambridge
  University Press},\ \bibinfo {year} {2010})\BibitemShut {NoStop}%
\bibitem [{\citenamefont {{Profumo}}(2017)}]{Profumo}%
  \BibitemOpen
  \bibfield  {author} {\bibinfo {author} {\bibfnamefont {S.}~\bibnamefont
  {{Profumo}}},\ }\href@noop {} {\emph {\bibinfo {title} {An Introduction to
  Particle Dark Matter}}}\ (\bibinfo  {publisher} {World Scientific},\ \bibinfo
  {year} {2017})\BibitemShut {NoStop}%
\bibitem [{\citenamefont {{Li}}\ \emph {et~al.}(2011)\citenamefont {{Li}},
  \citenamefont {{Li}}, \citenamefont {{Wang}},\ and\ \citenamefont
  {{Wang}}}]{Li2011}%
  \BibitemOpen
  \bibfield  {author} {\bibinfo {author} {\bibfnamefont {M.}~\bibnamefont
  {{Li}}}, \bibinfo {author} {\bibfnamefont {X.-D.}\ \bibnamefont {{Li}}},
  \bibinfo {author} {\bibfnamefont {S.}~\bibnamefont {{Wang}}}, \ and\ \bibinfo
  {author} {\bibfnamefont {Y.}~\bibnamefont {{Wang}}},\ }\href {\doibase
  10.1088/0253-6102/56/3/24} {\bibfield  {journal} {\bibinfo  {journal}
  {Communications in Theoretical Physics}\ }\textbf {\bibinfo {volume} {56}},\
  \bibinfo {pages} {525} (\bibinfo {year} {2011})},\ \Eprint
  {http://arxiv.org/abs/1103.5870} {arXiv:1103.5870} \BibitemShut {NoStop}%
\bibitem [{\citenamefont {{Del Popolo}}(2013)}]{DelPopolo2013}%
  \BibitemOpen
  \bibfield  {author} {\bibinfo {author} {\bibfnamefont {A.}~\bibnamefont {{Del
  Popolo}}},\ }in\ \href {\doibase 10.1063/1.4817029} {\emph {\bibinfo
  {booktitle} {AIP Conf. Proc.}}},\ Vol.\ \bibinfo {volume} {1548}\ (\bibinfo
  {year} {2013})\ pp.\ \bibinfo {pages} {2--63}\BibitemShut {NoStop}%
\bibitem [{\citenamefont {{Ade}}(2016)}]{Ade}%
  \BibitemOpen
  \bibfield  {author} {\bibinfo {author} {\bibfnamefont {P.~A. R. e.~a.}\
  \bibnamefont {{Ade}}},\ }\href@noop {} {\bibfield  {journal} {\bibinfo
  {journal} {Astron. Astrophys.}\ }\textbf {\bibinfo {volume} {594}},\ \bibinfo
  {pages} {A13} (\bibinfo {year} {2016})}\BibitemShut {NoStop}%
\bibitem [{\citenamefont {{de Blok}}(2010)}]{deBlok2010}%
  \BibitemOpen
  \bibfield  {author} {\bibinfo {author} {\bibfnamefont {W.~J.~G.}\
  \bibnamefont {{de Blok}}},\ }\href {\doibase 10.1155/2010/789293} {\bibfield
  {journal} {\bibinfo  {journal} {Advances in Astronomy}\ }\textbf {\bibinfo
  {volume} {2010}},\ \bibinfo {pages} {789293} (\bibinfo {year} {2010})},\
  \Eprint {http://arxiv.org/abs/0910.3538} {arXiv:0910.3538} \BibitemShut
  {NoStop}%
\bibitem [{\citenamefont {{Moore}}\ \emph {et~al.}(1999)\citenamefont
  {{Moore}}, \citenamefont {{Ghigna}}, \citenamefont {{Governato}},
  \citenamefont {{Lake}},\ and\ \citenamefont {{Quinn}}}]{Moore}%
  \BibitemOpen
  \bibfield  {author} {\bibinfo {author} {\bibfnamefont {B.}~\bibnamefont
  {{Moore}}}, \bibinfo {author} {\bibfnamefont {S.}~\bibnamefont {{Ghigna}}},
  \bibinfo {author} {\bibfnamefont {F.}~\bibnamefont {{Governato}}}, \bibinfo
  {author} {\bibfnamefont {G.}~\bibnamefont {{Lake}}}, \ and\ \bibinfo {author}
  {\bibfnamefont {T.}~\bibnamefont {{Quinn}}},\ }\href@noop {} {\bibfield
  {journal} {\bibinfo  {journal} {Astrophys. J.}\ }\textbf {\bibinfo {volume}
  {524}},\ \bibinfo {pages} {L19} (\bibinfo {year} {1999})}\BibitemShut
  {NoStop}%
\bibitem [{\citenamefont {{McGaugh}}(2004)}]{McGaugh1}%
  \BibitemOpen
  \bibfield  {author} {\bibinfo {author} {\bibfnamefont {S.~S.}\ \bibnamefont
  {{McGaugh}}},\ }\href@noop {} {\bibfield  {journal} {\bibinfo  {journal}
  {Astrophys. J.}\ }\textbf {\bibinfo {volume} {609}},\ \bibinfo {pages} {652}
  (\bibinfo {year} {2004})}\BibitemShut {NoStop}%
\bibitem [{\citenamefont {{McGaugh}}\ \emph {et~al.}(2016)\citenamefont
  {{McGaugh}}, \citenamefont {{Lelli}},\ and\ \citenamefont
  {{Schombert}}}]{McGaugh2}%
  \BibitemOpen
  \bibfield  {author} {\bibinfo {author} {\bibfnamefont {S.~S.}\ \bibnamefont
  {{McGaugh}}}, \bibinfo {author} {\bibfnamefont {F.}~\bibnamefont {{Lelli}}},
  \ and\ \bibinfo {author} {\bibfnamefont {J.~M.}\ \bibnamefont
  {{Schombert}}},\ }\href@noop {} {\bibfield  {journal} {\bibinfo  {journal}
  {Phys. Rev. Lett.}\ }\textbf {\bibinfo {volume} {117}},\ \bibinfo {pages}
  {201101} (\bibinfo {year} {2016})}\BibitemShut {NoStop}%
\bibitem [{\citenamefont {{Aprile}}(2017)}]{Aprile}%
  \BibitemOpen
  \bibfield  {author} {\bibinfo {author} {\bibfnamefont {E.~e.~a.}\
  \bibnamefont {{Aprile}}},\ }\href@noop {} {\bibfield  {journal} {\bibinfo
  {journal} {Phys. Rev. Lett.}\ }\textbf {\bibinfo {volume} {119}},\ \bibinfo
  {pages} {181301} (\bibinfo {year} {2017})}\BibitemShut {NoStop}%
\bibitem [{\citenamefont {{Abecrcrombie}}(2020)}]{Abecrcrombie}%
  \BibitemOpen
  \bibfield  {author} {\bibinfo {author} {\bibfnamefont {D.~e.~a.}\
  \bibnamefont {{Abecrcrombie}}},\ }\href@noop {} {\bibfield  {journal}
  {\bibinfo  {journal} {Physics of Dark Universe}\ }\textbf {\bibinfo {volume}
  {27}},\ \bibinfo {pages} {100371} (\bibinfo {year} {2020})}\BibitemShut
  {NoStop}%
\bibitem [{\citenamefont {{Ackermann}}(2015)}]{Ackermann}%
  \BibitemOpen
  \bibfield  {author} {\bibinfo {author} {\bibfnamefont {M.~e.~a.}\
  \bibnamefont {{Ackermann}}},\ }\href@noop {} {\bibfield  {journal} {\bibinfo
  {journal} {Phys. Rev. Lett.}\ }\textbf {\bibinfo {volume} {115}},\ \bibinfo
  {pages} {231301} (\bibinfo {year} {2015})}\BibitemShut {NoStop}%
\bibitem [{\citenamefont {{Chan}}\ and\ \citenamefont
  {{Leung}}(2017)}]{Chan2017}%
  \BibitemOpen
  \bibfield  {author} {\bibinfo {author} {\bibfnamefont {M.~H.}\ \bibnamefont
  {{Chan}}}\ and\ \bibinfo {author} {\bibfnamefont {C.~H.}\ \bibnamefont
  {{Leung}}},\ }\href {\doibase 10.1038/s41598-017-14950-4} {\bibfield
  {journal} {\bibinfo  {journal} {Scientific Reports}\ }\textbf {\bibinfo
  {volume} {7}},\ \bibinfo {pages} {14895} (\bibinfo {year} {2017})},\ \Eprint
  {http://arxiv.org/abs/1710.08123} {arXiv:1710.08123} \BibitemShut {NoStop}%
\bibitem [{\citenamefont {{Chan}}\ \emph {et~al.}(2019)\citenamefont {{Chan}},
  \citenamefont {{Cui}}, \citenamefont {{Liu}},\ and\ \citenamefont
  {{Leung}}}]{Chan2019}%
  \BibitemOpen
  \bibfield  {author} {\bibinfo {author} {\bibfnamefont {M.~H.}\ \bibnamefont
  {{Chan}}}, \bibinfo {author} {\bibfnamefont {L.}~\bibnamefont {{Cui}}},
  \bibinfo {author} {\bibfnamefont {J.}~\bibnamefont {{Liu}}}, \ and\ \bibinfo
  {author} {\bibfnamefont {C.~S.}\ \bibnamefont {{Leung}}},\ }\href {\doibase
  10.3847/1538-4357/aafe0b} {\bibfield  {journal} {\bibinfo  {journal} {\apj}\
  }\textbf {\bibinfo {volume} {872}},\ \bibinfo {pages} {177} (\bibinfo {year}
  {2019})},\ \Eprint {http://arxiv.org/abs/1901.04638} {arXiv:1901.04638}
  \BibitemShut {NoStop}%
\bibitem [{\citenamefont {{Chan}}\ and\ \citenamefont {{Lee}}(2020)}]{Chan3}%
  \BibitemOpen
  \bibfield  {author} {\bibinfo {author} {\bibfnamefont {M.~H.}\ \bibnamefont
  {{Chan}}}\ and\ \bibinfo {author} {\bibfnamefont {C.~M.}\ \bibnamefont
  {{Lee}}},\ }\href@noop {} {\bibfield  {journal} {\bibinfo  {journal} {Phys.
  Rev. D}\ } (\bibinfo {year} {2020})}\BibitemShut {NoStop}%
\bibitem [{\citenamefont {{Weinberg}}(1989)}]{Weinberg1989}%
  \BibitemOpen
  \bibfield  {author} {\bibinfo {author} {\bibfnamefont {S.}~\bibnamefont
  {{Weinberg}}},\ }\href {\doibase 10.1103/RevModPhys.61.1} {\bibfield
  {journal} {\bibinfo  {journal} {Reviews of Modern Physics}\ }\textbf
  {\bibinfo {volume} {61}},\ \bibinfo {pages} {1} (\bibinfo {year}
  {1989})}\BibitemShut {NoStop}%
\bibitem [{\citenamefont {{Velten}}\ \emph {et~al.}(2014)\citenamefont
  {{Velten}}, \citenamefont {{vom Marttens}},\ and\ \citenamefont
  {{Zimdahl}}}]{Velten2014}%
  \BibitemOpen
  \bibfield  {author} {\bibinfo {author} {\bibfnamefont {H.~E.~S.}\
  \bibnamefont {{Velten}}}, \bibinfo {author} {\bibfnamefont {R.~F.}\
  \bibnamefont {{vom Marttens}}}, \ and\ \bibinfo {author} {\bibfnamefont
  {W.}~\bibnamefont {{Zimdahl}}},\ }\href {\doibase
  10.1140/epjc/s10052-014-3160-4} {\bibfield  {journal} {\bibinfo  {journal}
  {European Physical Journal C}\ }\textbf {\bibinfo {volume} {74}},\ \bibinfo
  {pages} {3160} (\bibinfo {year} {2014})},\ \Eprint
  {http://arxiv.org/abs/1410.2509} {arXiv:1410.2509} \BibitemShut {NoStop}%
\bibitem [{\citenamefont {{Copeland}}\ \emph {et~al.}(2006)\citenamefont
  {{Copeland}}, \citenamefont {{Sami}},\ and\ \citenamefont
  {{Tsujikawa}}}]{Copeland2006}%
  \BibitemOpen
  \bibfield  {author} {\bibinfo {author} {\bibfnamefont {E.~J.}\ \bibnamefont
  {{Copeland}}}, \bibinfo {author} {\bibfnamefont {M.}~\bibnamefont {{Sami}}},
  \ and\ \bibinfo {author} {\bibfnamefont {S.}~\bibnamefont {{Tsujikawa}}},\
  }\href {\doibase 10.1142/S021827180600942X} {\bibfield  {journal} {\bibinfo
  {journal} {International Journal of Modern Physics D}\ }\textbf {\bibinfo
  {volume} {15}},\ \bibinfo {pages} {1753} (\bibinfo {year} {2006})},\ \Eprint
  {http://arxiv.org/abs/arXiv:hep-th/0603057} {arXiv:hep-th/0603057}
  \BibitemShut {NoStop}%
\bibitem [{\citenamefont {{Horndeski}}(1974)}]{Horndeski1974}%
  \BibitemOpen
  \bibfield  {author} {\bibinfo {author} {\bibfnamefont {G.~W.}\ \bibnamefont
  {{Horndeski}}},\ }\href {\doibase 10.1007/BF01807638} {\bibfield  {journal}
  {\bibinfo  {journal} {International Journal of Theoretical Physics}\ }\textbf
  {\bibinfo {volume} {10}},\ \bibinfo {pages} {363} (\bibinfo {year}
  {1974})}\BibitemShut {NoStop}%
\bibitem [{\citenamefont {{Milgrom}}(1983)}]{Milgrom1983}%
  \BibitemOpen
  \bibfield  {author} {\bibinfo {author} {\bibfnamefont {M.}~\bibnamefont
  {{Milgrom}}},\ }\href {\doibase 10.1086/161130} {\bibfield  {journal}
  {\bibinfo  {journal} {\apj}\ }\textbf {\bibinfo {volume} {270}},\ \bibinfo
  {pages} {365} (\bibinfo {year} {1983})}\BibitemShut {NoStop}%
\bibitem [{\citenamefont {{Zwiebach}}(1985)}]{Zwiebach1985}%
  \BibitemOpen
  \bibfield  {author} {\bibinfo {author} {\bibfnamefont {B.}~\bibnamefont
  {{Zwiebach}}},\ }\href {\doibase 10.1016/0370-2693(85)91616-8} {\bibfield
  {journal} {\bibinfo  {journal} {Physics Letters B}\ }\textbf {\bibinfo
  {volume} {156}},\ \bibinfo {pages} {315} (\bibinfo {year}
  {1985})}\BibitemShut {NoStop}%
\bibitem [{\citenamefont {{Moffat}}(2006)}]{Moffat2006}%
  \BibitemOpen
  \bibfield  {author} {\bibinfo {author} {\bibfnamefont {J.~W.}\ \bibnamefont
  {{Moffat}}},\ }\href {\doibase 10.1088/1475-7516/2006/03/004} {\bibfield
  {journal} {\bibinfo  {journal} {\jcap}\ }\textbf {\bibinfo {volume} {3}},\
  \bibinfo {pages} {004} (\bibinfo {year} {2006})},\ \Eprint
  {http://arxiv.org/abs/gr-qc/0506021} {gr-qc/0506021} \BibitemShut {NoStop}%
\bibitem [{\citenamefont {{Nojiri}}\ \emph {et~al.}(2005)\citenamefont
  {{Nojiri}}, \citenamefont {{Odintsov}},\ and\ \citenamefont
  {{Sasaki}}}]{Nojiri2005}%
  \BibitemOpen
  \bibfield  {author} {\bibinfo {author} {\bibfnamefont {S.}~\bibnamefont
  {{Nojiri}}}, \bibinfo {author} {\bibfnamefont {S.~D.}\ \bibnamefont
  {{Odintsov}}}, \ and\ \bibinfo {author} {\bibfnamefont {M.}~\bibnamefont
  {{Sasaki}}},\ }\href {\doibase 10.1103/PhysRevD.71.123509} {\bibfield
  {journal} {\bibinfo  {journal} {\prd}\ }\textbf {\bibinfo {volume} {71}},\
  \bibinfo {pages} {123509} (\bibinfo {year} {2005})},\ \Eprint
  {http://arxiv.org/abs/hep-th/0504052} {hep-th/0504052} \BibitemShut {NoStop}%
\bibitem [{\citenamefont {{Bekenstein}}(2010)}]{Bekenstein2010}%
  \BibitemOpen
  \bibfield  {author} {\bibinfo {author} {\bibfnamefont {J.~D.}\ \bibnamefont
  {{Bekenstein}}},\ }\enquote {\bibinfo {title} {{Modified gravity as an
  alternative to dark matter}},}\ in\ \href@noop {} {\emph {\bibinfo
  {booktitle} {Particle Dark Matter : Observations, Models and Searches}}},\
  \bibinfo {editor} {edited by\ \bibinfo {editor} {\bibfnamefont
  {G.}~\bibnamefont {{Bertone}}}}\ (\bibinfo  {publisher} {Cambridge University
  Press},\ \bibinfo {year} {2010})\ p.~\bibinfo {pages} {99}\BibitemShut
  {NoStop}%
\bibitem [{\citenamefont {{De Felice}}\ and\ \citenamefont
  {{Tsujikawa}}(2010)}]{DeFelice2010}%
  \BibitemOpen
  \bibfield  {author} {\bibinfo {author} {\bibfnamefont {A.}~\bibnamefont {{De
  Felice}}}\ and\ \bibinfo {author} {\bibfnamefont {S.}~\bibnamefont
  {{Tsujikawa}}},\ }\href {\doibase 10.12942/lrr-2010-3} {\bibfield  {journal}
  {\bibinfo  {journal} {Living Reviews in Relativity}\ }\textbf {\bibinfo
  {volume} {13}},\ \bibinfo {pages} {3} (\bibinfo {year} {2010})},\ \Eprint
  {http://arxiv.org/abs/1002.4928} {arXiv:1002.4928} \BibitemShut {NoStop}%
\bibitem [{\citenamefont {{Linder}}(2010)}]{Linder2010}%
  \BibitemOpen
  \bibfield  {author} {\bibinfo {author} {\bibfnamefont {E.~V.}\ \bibnamefont
  {{Linder}}},\ }\href {\doibase 10.1103/PhysRevD.81.127301} {\bibfield
  {journal} {\bibinfo  {journal} {\prd}\ }\textbf {\bibinfo {volume} {81}},\
  \bibinfo {pages} {127301} (\bibinfo {year} {2010})},\ \Eprint
  {http://arxiv.org/abs/1005.3039} {arXiv:1005.3039} \BibitemShut {NoStop}%
\bibitem [{\citenamefont {{Milgrom}}(2014)}]{Milgrom2014}%
  \BibitemOpen
  \bibfield  {author} {\bibinfo {author} {\bibfnamefont {M.}~\bibnamefont
  {{Milgrom}}},\ }\href {\doibase 10.1103/PhysRevD.89.024027} {\bibfield
  {journal} {\bibinfo  {journal} {\prd}\ }\textbf {\bibinfo {volume} {89}},\
  \bibinfo {pages} {024027} (\bibinfo {year} {2014})},\ \Eprint
  {http://arxiv.org/abs/1308.5388} {arXiv:1308.5388} \BibitemShut {NoStop}%
\bibitem [{\citenamefont {{Lovelock}}(1971)}]{Lovelock1971}%
  \BibitemOpen
  \bibfield  {author} {\bibinfo {author} {\bibfnamefont {D.}~\bibnamefont
  {{Lovelock}}},\ }\href {\doibase 10.1063/1.1665613} {\bibfield  {journal}
  {\bibinfo  {journal} {Journal of Mathematical Physics}\ }\textbf {\bibinfo
  {volume} {12}},\ \bibinfo {pages} {498} (\bibinfo {year} {1971})}\BibitemShut
  {NoStop}%
\bibitem [{\citenamefont {{Ho{\v r}ava}}(2009)}]{Horava2009}%
  \BibitemOpen
  \bibfield  {author} {\bibinfo {author} {\bibfnamefont {P.}~\bibnamefont
  {{Ho{\v r}ava}}},\ }\href {\doibase 10.1103/PhysRevD.79.084008} {\bibfield
  {journal} {\bibinfo  {journal} {\prd}\ }\textbf {\bibinfo {volume} {79}},\
  \bibinfo {pages} {084008} (\bibinfo {year} {2009})},\ \Eprint
  {http://arxiv.org/abs/0901.3775} {arXiv:0901.3775} \BibitemShut {NoStop}%
\bibitem [{\citenamefont {{Rodr{\'{\i}}guez}}\ and\ \citenamefont
  {{Navarro}}(2017)}]{Rodriguez2017}%
  \BibitemOpen
  \bibfield  {author} {\bibinfo {author} {\bibfnamefont {Y.}~\bibnamefont
  {{Rodr{\'{\i}}guez}}}\ and\ \bibinfo {author} {\bibfnamefont {A.~A.}\
  \bibnamefont {{Navarro}}},\ }in\ \href {\doibase
  10.1088/1742-6596/831/1/012004} {\emph {\bibinfo {booktitle} {Journal of
  Physics Conference Series}}},\ \bibinfo {series} {Journal of Physics
  Conference Series}, Vol.\ \bibinfo {volume} {831}\ (\bibinfo {year} {2017})\
  p.\ \bibinfo {pages} {012004},\ \Eprint {http://arxiv.org/abs/1703.01884}
  {arXiv:1703.01884} \BibitemShut {NoStop}%
\bibitem [{\citenamefont {{Deffayet}}\ \emph {et~al.}(2010)\citenamefont
  {{Deffayet}}, \citenamefont {{Pujol{\`a}s}}, \citenamefont {{Sawicki}},\ and\
  \citenamefont {{Vikman}}}]{Deffayet2010}%
  \BibitemOpen
  \bibfield  {author} {\bibinfo {author} {\bibfnamefont {C.}~\bibnamefont
  {{Deffayet}}}, \bibinfo {author} {\bibfnamefont {O.}~\bibnamefont
  {{Pujol{\`a}s}}}, \bibinfo {author} {\bibfnamefont {I.}~\bibnamefont
  {{Sawicki}}}, \ and\ \bibinfo {author} {\bibfnamefont {A.}~\bibnamefont
  {{Vikman}}},\ }\href {\doibase 10.1088/1475-7516/2010/10/026} {\bibfield
  {journal} {\bibinfo  {journal} {\jcap}\ }\textbf {\bibinfo {volume} {10}},\
  \bibinfo {pages} {026} (\bibinfo {year} {2010})},\ \Eprint
  {http://arxiv.org/abs/1008.0048} {arXiv:1008.0048} \BibitemShut {NoStop}%
\bibitem [{\citenamefont {{Verlinde}}(2017)}]{Verlinde}%
  \BibitemOpen
  \bibfield  {author} {\bibinfo {author} {\bibfnamefont {E.~P.}\ \bibnamefont
  {{Verlinde}}},\ }\href@noop {} {\bibfield  {journal} {\bibinfo  {journal}
  {SciPost Physics}\ }\textbf {\bibinfo {volume} {2}},\ \bibinfo {pages} {016}
  (\bibinfo {year} {2017})}\BibitemShut {NoStop}%
\bibitem [{\citenamefont {{Buchdahl}}(1970)}]{Buchdahl1970}%
  \BibitemOpen
  \bibfield  {author} {\bibinfo {author} {\bibfnamefont {H.~A.}\ \bibnamefont
  {{Buchdahl}}},\ }\href@noop {} {\bibfield  {journal} {\bibinfo  {journal}
  {\mnras}\ }\textbf {\bibinfo {volume} {150}},\ \bibinfo {pages} {1} (\bibinfo
  {year} {1970})}\BibitemShut {NoStop}%
\bibitem [{\citenamefont {{Bhattacharya}}\ \emph {et~al.}(2017)\citenamefont
  {{Bhattacharya}}, \citenamefont {{Dialektopoulos}}, \citenamefont {{Enea
  Romano}}, \citenamefont {{Skordis}},\ and\ \citenamefont
  {{Tomaras}}}]{Bhattacharya2017}%
  \BibitemOpen
  \bibfield  {author} {\bibinfo {author} {\bibfnamefont {S.}~\bibnamefont
  {{Bhattacharya}}}, \bibinfo {author} {\bibfnamefont {K.~F.}\ \bibnamefont
  {{Dialektopoulos}}}, \bibinfo {author} {\bibfnamefont {A.}~\bibnamefont
  {{Enea Romano}}}, \bibinfo {author} {\bibfnamefont {C.}~\bibnamefont
  {{Skordis}}}, \ and\ \bibinfo {author} {\bibfnamefont {T.~N.}\ \bibnamefont
  {{Tomaras}}},\ }\href {\doibase 10.1088/1475-7516/2017/07/018} {\bibfield
  {journal} {\bibinfo  {journal} {\jcap}\ }\textbf {\bibinfo {volume} {7}},\
  \bibinfo {pages} {018} (\bibinfo {year} {2017})},\ \Eprint
  {http://arxiv.org/abs/1611.05055} {arXiv:1611.05055} \BibitemShut {NoStop}%
\bibitem [{\citenamefont {{Lopes}}\ \emph {et~al.}(2018)\citenamefont
  {{Lopes}}, \citenamefont {{Voivodic}}, \citenamefont {{Abramo}},\ and\
  \citenamefont {{Sodr{\'e}}}}]{Lopes2018}%
  \BibitemOpen
  \bibfield  {author} {\bibinfo {author} {\bibfnamefont {R.~C.~C.}\
  \bibnamefont {{Lopes}}}, \bibinfo {author} {\bibfnamefont {R.}~\bibnamefont
  {{Voivodic}}}, \bibinfo {author} {\bibfnamefont {L.~R.}\ \bibnamefont
  {{Abramo}}}, \ and\ \bibinfo {author} {\bibfnamefont {J.}~\bibnamefont
  {{Sodr{\'e}}}, \bibfnamefont {Laerte}},\ }\href {\doibase
  10.1088/1475-7516/2018/09/010} {\bibfield  {journal} {\bibinfo  {journal}
  {\jcap}\ }\textbf {\bibinfo {volume} {2018}},\ \bibinfo {pages} {010}
  (\bibinfo {year} {2018})},\ \Eprint {http://arxiv.org/abs/1805.09918}
  {arXiv:1805.09918} \BibitemShut {NoStop}%
\bibitem [{\citenamefont {{Lopes}}\ \emph {et~al.}(2019)\citenamefont
  {{Lopes}}, \citenamefont {{Voivodic}}, \citenamefont {{Abramo}},\ and\
  \citenamefont {{Sodr{\'e}}}}]{Lopes2019}%
  \BibitemOpen
  \bibfield  {author} {\bibinfo {author} {\bibfnamefont {R.~C.~C.}\
  \bibnamefont {{Lopes}}}, \bibinfo {author} {\bibfnamefont {R.}~\bibnamefont
  {{Voivodic}}}, \bibinfo {author} {\bibfnamefont {L.~R.}\ \bibnamefont
  {{Abramo}}}, \ and\ \bibinfo {author} {\bibfnamefont {J.}~\bibnamefont
  {{Sodr{\'e}}}, \bibfnamefont {Laerte}},\ }\href {\doibase
  10.1088/1475-7516/2019/07/026} {\bibfield  {journal} {\bibinfo  {journal}
  {\jcap}\ }\textbf {\bibinfo {volume} {2019}},\ \bibinfo {pages} {026}
  (\bibinfo {year} {2019})},\ \Eprint {http://arxiv.org/abs/1809.10321}
  {arXiv:1809.10321} \BibitemShut {NoStop}%
\bibitem [{\citenamefont {{Pavlidou}}\ and\ \citenamefont
  {{Tomaras}}(2014)}]{Pavlidou2014}%
  \BibitemOpen
  \bibfield  {author} {\bibinfo {author} {\bibfnamefont {V.}~\bibnamefont
  {{Pavlidou}}}\ and\ \bibinfo {author} {\bibfnamefont {T.~N.}\ \bibnamefont
  {{Tomaras}}},\ }\href {\doibase 10.1088/1475-7516/2014/09/020} {\bibfield
  {journal} {\bibinfo  {journal} {\jcap}\ }\textbf {\bibinfo {volume} {2014}},\
  \bibinfo {pages} {020} (\bibinfo {year} {2014})},\ \Eprint
  {http://arxiv.org/abs/1310.1920} {arXiv:1310.1920} \BibitemShut {NoStop}%
\bibitem [{\citenamefont {{Pavlidou}}\ \emph {et~al.}(2014)\citenamefont
  {{Pavlidou}}, \citenamefont {{Tetradis}},\ and\ \citenamefont
  {{Tomaras}}}]{Pavlidou2014a}%
  \BibitemOpen
  \bibfield  {author} {\bibinfo {author} {\bibfnamefont {V.}~\bibnamefont
  {{Pavlidou}}}, \bibinfo {author} {\bibfnamefont {N.}~\bibnamefont
  {{Tetradis}}}, \ and\ \bibinfo {author} {\bibfnamefont {T.~N.}\ \bibnamefont
  {{Tomaras}}},\ }\href {\doibase 10.1088/1475-7516/2014/05/017} {\bibfield
  {journal} {\bibinfo  {journal} {\jcap}\ }\textbf {\bibinfo {volume} {2014}},\
  \bibinfo {pages} {017} (\bibinfo {year} {2014})},\ \Eprint
  {http://arxiv.org/abs/1401.3742} {arXiv:1401.3742} \BibitemShut {NoStop}%
\bibitem [{\citenamefont {{Faraoni}}\ \emph {et~al.}(2015)\citenamefont
  {{Faraoni}}, \citenamefont {{Lapierre-L{\'e}onard}},\ and\ \citenamefont
  {{Prain}}}]{Faraoni2015}%
  \BibitemOpen
  \bibfield  {author} {\bibinfo {author} {\bibfnamefont {V.}~\bibnamefont
  {{Faraoni}}}, \bibinfo {author} {\bibfnamefont {M.}~\bibnamefont
  {{Lapierre-L{\'e}onard}}}, \ and\ \bibinfo {author} {\bibfnamefont
  {A.}~\bibnamefont {{Prain}}},\ }\href {\doibase
  10.1088/1475-7516/2015/10/013} {\bibfield  {journal} {\bibinfo  {journal}
  {\jcap}\ }\textbf {\bibinfo {volume} {2015}},\ \bibinfo {pages} {013}
  (\bibinfo {year} {2015})},\ \Eprint {http://arxiv.org/abs/1508.01725}
  {arXiv:1508.01725} \BibitemShut {NoStop}%
\bibitem [{\citenamefont {{Del Popolo}}\ \emph
  {et~al.}(2013{\natexlab{a}})\citenamefont {{Del Popolo}}, \citenamefont
  {{Pace}},\ and\ \citenamefont {{Lima}}}]{DelPopolo2013a}%
  \BibitemOpen
  \bibfield  {author} {\bibinfo {author} {\bibfnamefont {A.}~\bibnamefont {{Del
  Popolo}}}, \bibinfo {author} {\bibfnamefont {F.}~\bibnamefont {{Pace}}}, \
  and\ \bibinfo {author} {\bibfnamefont {J.~A.~S.}\ \bibnamefont {{Lima}}},\
  }\href {\doibase 10.1142/S0218271813500387} {\bibfield  {journal} {\bibinfo
  {journal} {International Journal of Modern Physics D}\ }\textbf {\bibinfo
  {volume} {22}},\ \bibinfo {pages} {50038} (\bibinfo {year}
  {2013}{\natexlab{a}})},\ \Eprint {http://arxiv.org/abs/1207.5789}
  {arXiv:1207.5789} \BibitemShut {NoStop}%
\bibitem [{\citenamefont {{Del Popolo}}\ \emph
  {et~al.}(2013{\natexlab{b}})\citenamefont {{Del Popolo}}, \citenamefont
  {{Pace}},\ and\ \citenamefont {{Lima}}}]{DelPopolo2013b}%
  \BibitemOpen
  \bibfield  {author} {\bibinfo {author} {\bibfnamefont {A.}~\bibnamefont {{Del
  Popolo}}}, \bibinfo {author} {\bibfnamefont {F.}~\bibnamefont {{Pace}}}, \
  and\ \bibinfo {author} {\bibfnamefont {J.~A.~S.}\ \bibnamefont {{Lima}}},\
  }\href {\doibase 10.1093/mnras/sts669} {\bibfield  {journal} {\bibinfo
  {journal} {\mnras}\ }\textbf {\bibinfo {volume} {430}},\ \bibinfo {pages}
  {628} (\bibinfo {year} {2013}{\natexlab{b}})},\ \Eprint
  {http://arxiv.org/abs/1212.5092} {arXiv:1212.5092} \BibitemShut {NoStop}%
\bibitem [{\citenamefont {{Pace}}\ \emph
  {et~al.}(2014{\natexlab{a}})\citenamefont {{Pace}}, \citenamefont
  {{Batista}},\ and\ \citenamefont {{Del Popolo}}}]{Pace2014}%
  \BibitemOpen
  \bibfield  {author} {\bibinfo {author} {\bibfnamefont {F.}~\bibnamefont
  {{Pace}}}, \bibinfo {author} {\bibfnamefont {R.~C.}\ \bibnamefont
  {{Batista}}}, \ and\ \bibinfo {author} {\bibfnamefont {A.}~\bibnamefont {{Del
  Popolo}}},\ }\href {\doibase 10.1093/mnras/stu1782} {\bibfield  {journal}
  {\bibinfo  {journal} {\mnras}\ }\textbf {\bibinfo {volume} {445}},\ \bibinfo
  {pages} {648} (\bibinfo {year} {2014}{\natexlab{a}})},\ \Eprint
  {http://arxiv.org/abs/1406.1448} {arXiv:1406.1448} \BibitemShut {NoStop}%
\bibitem [{\citenamefont {{Mehrabi}}\ \emph {et~al.}(2017)\citenamefont
  {{Mehrabi}}, \citenamefont {{Pace}}, \citenamefont {{Malekjani}},\ and\
  \citenamefont {{Del Popolo}}}]{Mehrabi2017}%
  \BibitemOpen
  \bibfield  {author} {\bibinfo {author} {\bibfnamefont {A.}~\bibnamefont
  {{Mehrabi}}}, \bibinfo {author} {\bibfnamefont {F.}~\bibnamefont {{Pace}}},
  \bibinfo {author} {\bibfnamefont {M.}~\bibnamefont {{Malekjani}}}, \ and\
  \bibinfo {author} {\bibfnamefont {A.}~\bibnamefont {{Del Popolo}}},\ }\href
  {\doibase 10.1093/mnras/stw2927} {\bibfield  {journal} {\bibinfo  {journal}
  {\mnras}\ }\textbf {\bibinfo {volume} {465}},\ \bibinfo {pages} {2687}
  (\bibinfo {year} {2017})},\ \Eprint {http://arxiv.org/abs/1608.07961}
  {arXiv:1608.07961} \BibitemShut {NoStop}%
\bibitem [{\citenamefont {Pace}\ \emph {et~al.}(2019)\citenamefont {Pace},
  \citenamefont {Schimd}, \citenamefont {Mota},\ and\ \citenamefont
  {Popolo}}]{Pace2019}%
  \BibitemOpen
  \bibfield  {author} {\bibinfo {author} {\bibfnamefont {F.}~\bibnamefont
  {Pace}}, \bibinfo {author} {\bibfnamefont {C.}~\bibnamefont {Schimd}},
  \bibinfo {author} {\bibfnamefont {D.~F.}\ \bibnamefont {Mota}}, \ and\
  \bibinfo {author} {\bibfnamefont {A.~D.}\ \bibnamefont {Popolo}},\ }\href
  {\doibase 10.1088/1475-7516/2019/09/060} {\bibfield  {journal} {\bibinfo
  {journal} {Journal of Cosmology and Astroparticle Physics}\ }\textbf
  {\bibinfo {volume} {2019}},\ \bibinfo {pages} {060} (\bibinfo {year}
  {2019})}\BibitemShut {NoStop}%
\bibitem [{\citenamefont {{Del Popolo}}\ \emph {et~al.}(2020)\citenamefont
  {{Del Popolo}}, \citenamefont {{Chan}},\ and\ \citenamefont
  {{Mota}}}]{DelPopoloChan2020}%
  \BibitemOpen
  \bibfield  {author} {\bibinfo {author} {\bibfnamefont {A.}~\bibnamefont {{Del
  Popolo}}}, \bibinfo {author} {\bibfnamefont {M.~H.}\ \bibnamefont {{Chan}}},
  \ and\ \bibinfo {author} {\bibfnamefont {D.~F.}\ \bibnamefont {{Mota}}},\
  }\href {\doibase 10.1103/PhysRevD.101.083505} {\bibfield  {journal} {\bibinfo
   {journal} {\prd}\ }\textbf {\bibinfo {volume} {101}},\ \bibinfo {pages}
  {083505} (\bibinfo {year} {2020})}\BibitemShut {NoStop}%
\bibitem [{\citenamefont {{Del Popolo}}\ and\ \citenamefont
  {{Gambera}}(1999)}]{DelPopolo1999}%
  \BibitemOpen
  \bibfield  {author} {\bibinfo {author} {\bibfnamefont {A.}~\bibnamefont {{Del
  Popolo}}}\ and\ \bibinfo {author} {\bibfnamefont {M.}~\bibnamefont
  {{Gambera}}},\ }\href@noop {} {\bibfield  {journal} {\bibinfo  {journal}
  {\aap}\ }\textbf {\bibinfo {volume} {344}},\ \bibinfo {pages} {17} (\bibinfo
  {year} {1999})},\ \Eprint {http://arxiv.org/abs/arXiv:astro-ph/9806044}
  {arXiv:astro-ph/9806044} \BibitemShut {NoStop}%
\bibitem [{\citenamefont {{Fillmore}}\ and\ \citenamefont
  {{Goldreich}}(1984)}]{Fillmore1984}%
  \BibitemOpen
  \bibfield  {author} {\bibinfo {author} {\bibfnamefont {J.~A.}\ \bibnamefont
  {{Fillmore}}}\ and\ \bibinfo {author} {\bibfnamefont {P.}~\bibnamefont
  {{Goldreich}}},\ }\href {\doibase 10.1086/162070} {\bibfield  {journal}
  {\bibinfo  {journal} {\apj}\ }\textbf {\bibinfo {volume} {281}},\ \bibinfo
  {pages} {1} (\bibinfo {year} {1984})}\BibitemShut {NoStop}%
\bibitem [{\citenamefont {{Bertschinger}}(1985)}]{Bertschinger1985}%
  \BibitemOpen
  \bibfield  {author} {\bibinfo {author} {\bibfnamefont {E.}~\bibnamefont
  {{Bertschinger}}},\ }\href {\doibase 10.1086/191028} {\bibfield  {journal}
  {\bibinfo  {journal} {\apjs}\ }\textbf {\bibinfo {volume} {58}},\ \bibinfo
  {pages} {39} (\bibinfo {year} {1985})}\BibitemShut {NoStop}%
\bibitem [{\citenamefont {{Hoffman}}\ and\ \citenamefont
  {{Shaham}}(1985)}]{Hoffman1985}%
  \BibitemOpen
  \bibfield  {author} {\bibinfo {author} {\bibfnamefont {Y.}~\bibnamefont
  {{Hoffman}}}\ and\ \bibinfo {author} {\bibfnamefont {J.}~\bibnamefont
  {{Shaham}}},\ }\href {\doibase 10.1086/163498} {\bibfield  {journal}
  {\bibinfo  {journal} {\apj}\ }\textbf {\bibinfo {volume} {297}},\ \bibinfo
  {pages} {16} (\bibinfo {year} {1985})}\BibitemShut {NoStop}%
\bibitem [{\citenamefont {{Ryden}}\ and\ \citenamefont
  {{Gunn}}(1987)}]{Ryden1987}%
  \BibitemOpen
  \bibfield  {author} {\bibinfo {author} {\bibfnamefont {B.~S.}\ \bibnamefont
  {{Ryden}}}\ and\ \bibinfo {author} {\bibfnamefont {J.~E.}\ \bibnamefont
  {{Gunn}}},\ }\href {\doibase 10.1086/165349} {\bibfield  {journal} {\bibinfo
  {journal} {\apj}\ }\textbf {\bibinfo {volume} {318}},\ \bibinfo {pages} {15}
  (\bibinfo {year} {1987})}\BibitemShut {NoStop}%
\bibitem [{\citenamefont {{Subramanian}}\ \emph {et~al.}(2000)\citenamefont
  {{Subramanian}}, \citenamefont {{Cen}},\ and\ \citenamefont
  {{Ostriker}}}]{Subramanian2000}%
  \BibitemOpen
  \bibfield  {author} {\bibinfo {author} {\bibfnamefont {K.}~\bibnamefont
  {{Subramanian}}}, \bibinfo {author} {\bibfnamefont {R.}~\bibnamefont
  {{Cen}}}, \ and\ \bibinfo {author} {\bibfnamefont {J.~P.}\ \bibnamefont
  {{Ostriker}}},\ }\href {\doibase 10.1086/309152} {\bibfield  {journal}
  {\bibinfo  {journal} {\apj}\ }\textbf {\bibinfo {volume} {538}},\ \bibinfo
  {pages} {528} (\bibinfo {year} {2000})},\ \Eprint
  {http://arxiv.org/abs/arXiv:astro-ph/9909279} {arXiv:astro-ph/9909279}
  \BibitemShut {NoStop}%
\bibitem [{\citenamefont {{Ascasibar}}\ \emph {et~al.}(2004)\citenamefont
  {{Ascasibar}}, \citenamefont {{Yepes}}, \citenamefont {{Gottl{\"o}ber}},\
  and\ \citenamefont {{M{\"u}ller}}}]{Ascasibar2004}%
  \BibitemOpen
  \bibfield  {author} {\bibinfo {author} {\bibfnamefont {Y.}~\bibnamefont
  {{Ascasibar}}}, \bibinfo {author} {\bibfnamefont {G.}~\bibnamefont
  {{Yepes}}}, \bibinfo {author} {\bibfnamefont {S.}~\bibnamefont
  {{Gottl{\"o}ber}}}, \ and\ \bibinfo {author} {\bibfnamefont {V.}~\bibnamefont
  {{M{\"u}ller}}},\ }\href {\doibase 10.1111/j.1365-2966.2004.08005.x}
  {\bibfield  {journal} {\bibinfo  {journal} {\mnras}\ }\textbf {\bibinfo
  {volume} {352}},\ \bibinfo {pages} {1109} (\bibinfo {year} {2004})},\ \Eprint
  {http://arxiv.org/abs/arXiv:astro-ph/0312221} {arXiv:astro-ph/0312221}
  \BibitemShut {NoStop}%
\bibitem [{\citenamefont {{Williams}}\ \emph {et~al.}(2004)\citenamefont
  {{Williams}}, \citenamefont {{Babul}},\ and\ \citenamefont
  {{Dalcanton}}}]{Williams2004}%
  \BibitemOpen
  \bibfield  {author} {\bibinfo {author} {\bibfnamefont {L.~L.~R.}\
  \bibnamefont {{Williams}}}, \bibinfo {author} {\bibfnamefont
  {A.}~\bibnamefont {{Babul}}}, \ and\ \bibinfo {author} {\bibfnamefont
  {J.~J.}\ \bibnamefont {{Dalcanton}}},\ }\href {\doibase 10.1086/381722}
  {\bibfield  {journal} {\bibinfo  {journal} {\apj}\ }\textbf {\bibinfo
  {volume} {604}},\ \bibinfo {pages} {18} (\bibinfo {year} {2004})},\ \Eprint
  {http://arxiv.org/abs/arXiv:astro-ph/0312002} {arXiv:astro-ph/0312002}
  \BibitemShut {NoStop}%
\bibitem [{\citenamefont {{Gunn}}\ and\ \citenamefont
  {{Gott}}(1972)}]{Gunn1972}%
  \BibitemOpen
  \bibfield  {author} {\bibinfo {author} {\bibfnamefont {J.~E.}\ \bibnamefont
  {{Gunn}}}\ and\ \bibinfo {author} {\bibfnamefont {J.~R.}\ \bibnamefont
  {{Gott}}, \bibfnamefont {III}},\ }\href {\doibase 10.1086/151605} {\bibfield
  {journal} {\bibinfo  {journal} {\apj}\ }\textbf {\bibinfo {volume} {176}},\
  \bibinfo {pages} {1} (\bibinfo {year} {1972})}\BibitemShut {NoStop}%
\bibitem [{\citenamefont {{Gurevich}}\ and\ \citenamefont
  {{Zybin}}(1988{\natexlab{a}})}]{Gurevich1988a}%
  \BibitemOpen
  \bibfield  {author} {\bibinfo {author} {\bibfnamefont {A.~V.}\ \bibnamefont
  {{Gurevich}}}\ and\ \bibinfo {author} {\bibfnamefont {K.~P.}\ \bibnamefont
  {{Zybin}}},\ }\href@noop {} {\bibfield  {journal} {\bibinfo  {journal}
  {Zhurnal Eksperimental noi i Teoreticheskoi Fiziki}\ }\textbf {\bibinfo
  {volume} {94}},\ \bibinfo {pages} {3} (\bibinfo {year}
  {1988}{\natexlab{a}})}\BibitemShut {NoStop}%
\bibitem [{\citenamefont {{Gurevich}}\ and\ \citenamefont
  {{Zybin}}(1988{\natexlab{b}})}]{Gurevich1988b}%
  \BibitemOpen
  \bibfield  {author} {\bibinfo {author} {\bibfnamefont {A.~V.}\ \bibnamefont
  {{Gurevich}}}\ and\ \bibinfo {author} {\bibfnamefont {K.~P.}\ \bibnamefont
  {{Zybin}}},\ }\href@noop {} {\bibfield  {journal} {\bibinfo  {journal}
  {Zhurnal Eksperimental noi i Teoreticheskoi Fiziki}\ }\textbf {\bibinfo
  {volume} {94}},\ \bibinfo {pages} {5} (\bibinfo {year}
  {1988}{\natexlab{b}})}\BibitemShut {NoStop}%
\bibitem [{\citenamefont {{White}}\ and\ \citenamefont
  {{Zaritsky}}(1992)}]{White1992}%
  \BibitemOpen
  \bibfield  {author} {\bibinfo {author} {\bibfnamefont {S.~D.~M.}\
  \bibnamefont {{White}}}\ and\ \bibinfo {author} {\bibfnamefont
  {D.}~\bibnamefont {{Zaritsky}}},\ }\href {\doibase 10.1086/171552} {\bibfield
   {journal} {\bibinfo  {journal} {\apj}\ }\textbf {\bibinfo {volume} {394}},\
  \bibinfo {pages} {1} (\bibinfo {year} {1992})}\BibitemShut {NoStop}%
\bibitem [{\citenamefont {{Sikivie}}\ \emph {et~al.}(1997)\citenamefont
  {{Sikivie}}, \citenamefont {{Tkachev}},\ and\ \citenamefont
  {{Wang}}}]{Sikivie1997}%
  \BibitemOpen
  \bibfield  {author} {\bibinfo {author} {\bibfnamefont {P.}~\bibnamefont
  {{Sikivie}}}, \bibinfo {author} {\bibfnamefont {I.~I.}\ \bibnamefont
  {{Tkachev}}}, \ and\ \bibinfo {author} {\bibfnamefont {Y.}~\bibnamefont
  {{Wang}}},\ }\href {\doibase 10.1103/PhysRevD.56.1863} {\bibfield  {journal}
  {\bibinfo  {journal} {\prd}\ }\textbf {\bibinfo {volume} {56}},\ \bibinfo
  {pages} {1863} (\bibinfo {year} {1997})},\ \Eprint
  {http://arxiv.org/abs/arXiv:astro-ph/9609022} {arXiv:astro-ph/9609022}
  \BibitemShut {NoStop}%
\bibitem [{\citenamefont {{Nusser}}(2001)}]{Nusser2001}%
  \BibitemOpen
  \bibfield  {author} {\bibinfo {author} {\bibfnamefont {A.}~\bibnamefont
  {{Nusser}}},\ }\href {\doibase 10.1046/j.1365-8711.2001.04527.x} {\bibfield
  {journal} {\bibinfo  {journal} {\mnras}\ }\textbf {\bibinfo {volume} {325}},\
  \bibinfo {pages} {1397} (\bibinfo {year} {2001})},\ \Eprint
  {http://arxiv.org/abs/arXiv:astro-ph/0008217} {arXiv:astro-ph/0008217}
  \BibitemShut {NoStop}%
\bibitem [{\citenamefont {{Hiotelis}}(2002)}]{Hiotelis2002}%
  \BibitemOpen
  \bibfield  {author} {\bibinfo {author} {\bibfnamefont {N.}~\bibnamefont
  {{Hiotelis}}},\ }\href {\doibase 10.1051/0004-6361:20011620} {\bibfield
  {journal} {\bibinfo  {journal} {\aap}\ }\textbf {\bibinfo {volume} {382}},\
  \bibinfo {pages} {84} (\bibinfo {year} {2002})},\ \Eprint
  {http://arxiv.org/abs/arXiv:astro-ph/0111324} {arXiv:astro-ph/0111324}
  \BibitemShut {NoStop}%
\bibitem [{\citenamefont {{Le Delliou}}\ and\ \citenamefont
  {{Henriksen}}(2003)}]{LeDelliou2003}%
  \BibitemOpen
  \bibfield  {author} {\bibinfo {author} {\bibfnamefont {M.}~\bibnamefont {{Le
  Delliou}}}\ and\ \bibinfo {author} {\bibfnamefont {R.~N.}\ \bibnamefont
  {{Henriksen}}},\ }\href {\doibase 10.1051/0004-6361:20030922} {\bibfield
  {journal} {\bibinfo  {journal} {\aap}\ }\textbf {\bibinfo {volume} {408}},\
  \bibinfo {pages} {27} (\bibinfo {year} {2003})},\ \Eprint
  {http://arxiv.org/abs/arXiv:astro-ph/0307046} {arXiv:astro-ph/0307046}
  \BibitemShut {NoStop}%
\bibitem [{\citenamefont {{Zukin}}\ and\ \citenamefont
  {{Bertschinger}}(2010)}]{Zukin2010}%
  \BibitemOpen
  \bibfield  {author} {\bibinfo {author} {\bibfnamefont {P.}~\bibnamefont
  {{Zukin}}}\ and\ \bibinfo {author} {\bibfnamefont {E.}~\bibnamefont
  {{Bertschinger}}},\ }in\ \href@noop {} {\emph {\bibinfo {booktitle} {APS
  Meeting Abstracts}}}\ (\bibinfo {year} {2010})\ p.\ \bibinfo {pages}
  {13003}\BibitemShut {NoStop}%
\bibitem [{\citenamefont {{Antonuccio-Delogu}}\ and\ \citenamefont
  {{Colafrancesco}}(1994)}]{AntonuccioDelogu1994}%
  \BibitemOpen
  \bibfield  {author} {\bibinfo {author} {\bibfnamefont {V.}~\bibnamefont
  {{Antonuccio-Delogu}}}\ and\ \bibinfo {author} {\bibfnamefont
  {S.}~\bibnamefont {{Colafrancesco}}},\ }\href {\doibase 10.1086/174122}
  {\bibfield  {journal} {\bibinfo  {journal} {\apj}\ }\textbf {\bibinfo
  {volume} {427}},\ \bibinfo {pages} {72} (\bibinfo {year} {1994})}\BibitemShut
  {NoStop}%
\bibitem [{\citenamefont {{Del Popolo}}(2009)}]{Delpopolo2009}%
  \BibitemOpen
  \bibfield  {author} {\bibinfo {author} {\bibfnamefont {A.}~\bibnamefont {{Del
  Popolo}}},\ }\href {\doibase 10.1088/0004-637X/698/2/2093} {\bibfield
  {journal} {\bibinfo  {journal} {\apj}\ }\textbf {\bibinfo {volume} {698}},\
  \bibinfo {pages} {2093} (\bibinfo {year} {2009})},\ \Eprint
  {http://arxiv.org/abs/0906.4447} {arXiv:0906.4447} \BibitemShut {NoStop}%
\bibitem [{\citenamefont {{Hoffman}}(1986)}]{Hoffman1986}%
  \BibitemOpen
  \bibfield  {author} {\bibinfo {author} {\bibfnamefont {Y.}~\bibnamefont
  {{Hoffman}}},\ }\href {\doibase 10.1086/164520} {\bibfield  {journal}
  {\bibinfo  {journal} {\apj}\ }\textbf {\bibinfo {volume} {308}},\ \bibinfo
  {pages} {493} (\bibinfo {year} {1986})}\BibitemShut {NoStop}%
\bibitem [{\citenamefont {{Hoffman}}(1989)}]{Hoffman1989}%
  \BibitemOpen
  \bibfield  {author} {\bibinfo {author} {\bibfnamefont {Y.}~\bibnamefont
  {{Hoffman}}},\ }\href {\doibase 10.1086/167376} {\bibfield  {journal}
  {\bibinfo  {journal} {\apj}\ }\textbf {\bibinfo {volume} {340}},\ \bibinfo
  {pages} {69} (\bibinfo {year} {1989})}\BibitemShut {NoStop}%
\bibitem [{\citenamefont {{Zaroubi}}\ and\ \citenamefont
  {{Hoffman}}(1993)}]{Zaroubi1993}%
  \BibitemOpen
  \bibfield  {author} {\bibinfo {author} {\bibfnamefont {S.}~\bibnamefont
  {{Zaroubi}}}\ and\ \bibinfo {author} {\bibfnamefont {Y.}~\bibnamefont
  {{Hoffman}}},\ }\href {\doibase 10.1086/173246} {\bibfield  {journal}
  {\bibinfo  {journal} {\apj}\ }\textbf {\bibinfo {volume} {416}},\ \bibinfo
  {pages} {410} (\bibinfo {year} {1993})}\BibitemShut {NoStop}%
\bibitem [{\citenamefont {{Mota}}\ and\ \citenamefont {{van de
  Bruck}}(2004)}]{Mota2004}%
  \BibitemOpen
  \bibfield  {author} {\bibinfo {author} {\bibfnamefont {D.~F.}\ \bibnamefont
  {{Mota}}}\ and\ \bibinfo {author} {\bibfnamefont {C.}~\bibnamefont {{van de
  Bruck}}},\ }\href {\doibase 10.1051/0004-6361:20041090} {\bibfield  {journal}
  {\bibinfo  {journal} {\aap}\ }\textbf {\bibinfo {volume} {421}},\ \bibinfo
  {pages} {71} (\bibinfo {year} {2004})},\ \Eprint
  {http://arxiv.org/abs/arXiv:astro-ph/0401504} {arXiv:astro-ph/0401504}
  \BibitemShut {NoStop}%
\bibitem [{\citenamefont {{Nunes}}\ and\ \citenamefont
  {{Mota}}(2006)}]{Nunes2006}%
  \BibitemOpen
  \bibfield  {author} {\bibinfo {author} {\bibfnamefont {N.~J.}\ \bibnamefont
  {{Nunes}}}\ and\ \bibinfo {author} {\bibfnamefont {D.~F.}\ \bibnamefont
  {{Mota}}},\ }\href {\doibase 10.1111/j.1365-2966.2006.10166.x} {\bibfield
  {journal} {\bibinfo  {journal} {\mnras}\ }\textbf {\bibinfo {volume} {368}},\
  \bibinfo {pages} {751} (\bibinfo {year} {2006})},\ \Eprint
  {http://arxiv.org/abs/arXiv:astro-ph/0409481} {arXiv:astro-ph/0409481}
  \BibitemShut {NoStop}%
\bibitem [{\citenamefont {{Abramo}}\ \emph {et~al.}(2007)\citenamefont
  {{Abramo}}, \citenamefont {{Batista}}, \citenamefont {{Liberato}},\ and\
  \citenamefont {{Rosenfeld}}}]{Abramo2007}%
  \BibitemOpen
  \bibfield  {author} {\bibinfo {author} {\bibfnamefont {L.~R.}\ \bibnamefont
  {{Abramo}}}, \bibinfo {author} {\bibfnamefont {R.~C.}\ \bibnamefont
  {{Batista}}}, \bibinfo {author} {\bibfnamefont {L.}~\bibnamefont
  {{Liberato}}}, \ and\ \bibinfo {author} {\bibfnamefont {R.}~\bibnamefont
  {{Rosenfeld}}},\ }\href {\doibase 10.1088/1475-7516/2007/11/012} {\bibfield
  {journal} {\bibinfo  {journal} {Journal of Cosmology and Astro-Particle
  Physics}\ }\textbf {\bibinfo {volume} {11}},\ \bibinfo {pages} {12} (\bibinfo
  {year} {2007})},\ \Eprint {http://arxiv.org/abs/0707.2882} {arXiv:0707.2882}
  \BibitemShut {NoStop}%
\bibitem [{\citenamefont {{Abramo}}\ \emph {et~al.}(2008)\citenamefont
  {{Abramo}}, \citenamefont {{Batista}}, \citenamefont {{Liberato}},\ and\
  \citenamefont {{Rosenfeld}}}]{Abramo2008}%
  \BibitemOpen
  \bibfield  {author} {\bibinfo {author} {\bibfnamefont {L.~R.}\ \bibnamefont
  {{Abramo}}}, \bibinfo {author} {\bibfnamefont {R.~C.}\ \bibnamefont
  {{Batista}}}, \bibinfo {author} {\bibfnamefont {L.}~\bibnamefont
  {{Liberato}}}, \ and\ \bibinfo {author} {\bibfnamefont {R.}~\bibnamefont
  {{Rosenfeld}}},\ }\href {\doibase 10.1103/PhysRevD.77.067301} {\bibfield
  {journal} {\bibinfo  {journal} {\prd}\ }\textbf {\bibinfo {volume} {77}},\
  \bibinfo {pages} {067301} (\bibinfo {year} {2008})},\ \Eprint
  {http://arxiv.org/abs/0710.2368} {arXiv:0710.2368} \BibitemShut {NoStop}%
\bibitem [{\citenamefont {{Abramo}}\ \emph
  {et~al.}(2009{\natexlab{a}})\citenamefont {{Abramo}}, \citenamefont
  {{Batista}},\ and\ \citenamefont {{Rosenfeld}}}]{Abramo2009a}%
  \BibitemOpen
  \bibfield  {author} {\bibinfo {author} {\bibfnamefont {L.~R.}\ \bibnamefont
  {{Abramo}}}, \bibinfo {author} {\bibfnamefont {R.~C.}\ \bibnamefont
  {{Batista}}}, \ and\ \bibinfo {author} {\bibfnamefont {R.}~\bibnamefont
  {{Rosenfeld}}},\ }\href {\doibase 10.1088/1475-7516/2009/07/040} {\bibfield
  {journal} {\bibinfo  {journal} {Journal of Cosmology and Astro-Particle
  Physics}\ }\textbf {\bibinfo {volume} {7}},\ \bibinfo {pages} {40} (\bibinfo
  {year} {2009}{\natexlab{a}})},\ \Eprint {http://arxiv.org/abs/0902.3226}
  {arXiv:0902.3226} \BibitemShut {NoStop}%
\bibitem [{\citenamefont {{Abramo}}\ \emph
  {et~al.}(2009{\natexlab{b}})\citenamefont {{Abramo}}, \citenamefont
  {{Batista}}, \citenamefont {{Liberato}},\ and\ \citenamefont
  {{Rosenfeld}}}]{Abramo2009b}%
  \BibitemOpen
  \bibfield  {author} {\bibinfo {author} {\bibfnamefont {L.~R.}\ \bibnamefont
  {{Abramo}}}, \bibinfo {author} {\bibfnamefont {R.~C.}\ \bibnamefont
  {{Batista}}}, \bibinfo {author} {\bibfnamefont {L.}~\bibnamefont
  {{Liberato}}}, \ and\ \bibinfo {author} {\bibfnamefont {R.}~\bibnamefont
  {{Rosenfeld}}},\ }\href {\doibase 10.1103/PhysRevD.79.023516} {\bibfield
  {journal} {\bibinfo  {journal} {\prd}\ }\textbf {\bibinfo {volume} {79}},\
  \bibinfo {pages} {023516} (\bibinfo {year} {2009}{\natexlab{b}})},\ \Eprint
  {http://arxiv.org/abs/0806.3461} {arXiv:0806.3461} \BibitemShut {NoStop}%
\bibitem [{\citenamefont {{Creminelli}}\ \emph {et~al.}(2010)\citenamefont
  {{Creminelli}}, \citenamefont {{D'Amico}}, \citenamefont {{Nore{\~n}a}},
  \citenamefont {{Senatore}},\ and\ \citenamefont
  {{Vernizzi}}}]{Creminelli2010}%
  \BibitemOpen
  \bibfield  {author} {\bibinfo {author} {\bibfnamefont {P.}~\bibnamefont
  {{Creminelli}}}, \bibinfo {author} {\bibfnamefont {G.}~\bibnamefont
  {{D'Amico}}}, \bibinfo {author} {\bibfnamefont {J.}~\bibnamefont
  {{Nore{\~n}a}}}, \bibinfo {author} {\bibfnamefont {L.}~\bibnamefont
  {{Senatore}}}, \ and\ \bibinfo {author} {\bibfnamefont {F.}~\bibnamefont
  {{Vernizzi}}},\ }\href {\doibase 10.1088/1475-7516/2010/03/027} {\bibfield
  {journal} {\bibinfo  {journal} {\jcap}\ }\textbf {\bibinfo {volume} {3}},\
  \bibinfo {pages} {27} (\bibinfo {year} {2010})},\ \Eprint
  {http://arxiv.org/abs/0911.2701} {arXiv:0911.2701} \BibitemShut {NoStop}%
\bibitem [{\citenamefont {{Basse}}\ \emph {et~al.}(2011)\citenamefont
  {{Basse}}, \citenamefont {{Eggers Bj{\ae}lde}},\ and\ \citenamefont
  {{Wong}}}]{Basse2011}%
  \BibitemOpen
  \bibfield  {author} {\bibinfo {author} {\bibfnamefont {T.}~\bibnamefont
  {{Basse}}}, \bibinfo {author} {\bibfnamefont {O.}~\bibnamefont {{Eggers
  Bj{\ae}lde}}}, \ and\ \bibinfo {author} {\bibfnamefont {Y.~Y.~Y.}\
  \bibnamefont {{Wong}}},\ }\href {\doibase 10.1088/1475-7516/2011/10/038}
  {\bibfield  {journal} {\bibinfo  {journal} {\jcap}\ }\textbf {\bibinfo
  {volume} {10}},\ \bibinfo {pages} {38} (\bibinfo {year} {2011})},\ \Eprint
  {http://arxiv.org/abs/1009.0010} {arXiv:1009.0010} \BibitemShut {NoStop}%
\bibitem [{\citenamefont {{Batista}}\ and\ \citenamefont
  {{Pace}}(2013)}]{Batista2013}%
  \BibitemOpen
  \bibfield  {author} {\bibinfo {author} {\bibfnamefont {R.~C.}\ \bibnamefont
  {{Batista}}}\ and\ \bibinfo {author} {\bibfnamefont {F.}~\bibnamefont
  {{Pace}}},\ }\href {\doibase 10.1088/1475-7516/2013/06/044} {\bibfield
  {journal} {\bibinfo  {journal} {\jcap}\ }\textbf {\bibinfo {volume} {6}},\
  \bibinfo {pages} {44} (\bibinfo {year} {2013})},\ \Eprint
  {http://arxiv.org/abs/1303.0414} {arXiv:1303.0414} \BibitemShut {NoStop}%
\bibitem [{\citenamefont {{Pace}}\ \emph
  {et~al.}(2014{\natexlab{b}})\citenamefont {{Pace}}, \citenamefont
  {{Batista}},\ and\ \citenamefont {{Del Popolo}}}]{Pace2014b}%
  \BibitemOpen
  \bibfield  {author} {\bibinfo {author} {\bibfnamefont {F.}~\bibnamefont
  {{Pace}}}, \bibinfo {author} {\bibfnamefont {R.~C.}\ \bibnamefont
  {{Batista}}}, \ and\ \bibinfo {author} {\bibfnamefont {A.}~\bibnamefont {{Del
  Popolo}}},\ }\href {\doibase 10.1093/mnras/stu1782} {\bibfield  {journal}
  {\bibinfo  {journal} {\mnras}\ }\textbf {\bibinfo {volume} {445}},\ \bibinfo
  {pages} {648} (\bibinfo {year} {2014}{\natexlab{b}})},\ \Eprint
  {http://arxiv.org/abs/1406.1448} {arXiv:1406.1448} \BibitemShut {NoStop}%
\bibitem [{\citenamefont {{Del Popolo}}\ \emph
  {et~al.}(2013{\natexlab{c}})\citenamefont {{Del Popolo}}, \citenamefont
  {{Pace}}, \citenamefont {{Maydanyuk}}, \citenamefont {{Lima}},\ and\
  \citenamefont {{Jesus}}}]{DelPopolo2013c}%
  \BibitemOpen
  \bibfield  {author} {\bibinfo {author} {\bibfnamefont {A.}~\bibnamefont {{Del
  Popolo}}}, \bibinfo {author} {\bibfnamefont {F.}~\bibnamefont {{Pace}}},
  \bibinfo {author} {\bibfnamefont {S.~P.}\ \bibnamefont {{Maydanyuk}}},
  \bibinfo {author} {\bibfnamefont {J.~A.~S.}\ \bibnamefont {{Lima}}}, \ and\
  \bibinfo {author} {\bibfnamefont {J.~F.}\ \bibnamefont {{Jesus}}},\ }\href
  {\doibase 10.1103/PhysRevD.87.043527} {\bibfield  {journal} {\bibinfo
  {journal} {\prd}\ }\textbf {\bibinfo {volume} {87}},\ \bibinfo {pages}
  {043527} (\bibinfo {year} {2013}{\natexlab{c}})},\ \Eprint
  {http://arxiv.org/abs/1303.3628} {arXiv:1303.3628} \BibitemShut {NoStop}%
\bibitem [{\citenamefont {{Bernardeau}}(1994)}]{Bernardeau1994}%
  \BibitemOpen
  \bibfield  {author} {\bibinfo {author} {\bibfnamefont {F.}~\bibnamefont
  {{Bernardeau}}},\ }\href {\doibase 10.1086/174620} {\bibfield  {journal}
  {\bibinfo  {journal} {\apj}\ }\textbf {\bibinfo {volume} {433}},\ \bibinfo
  {pages} {1} (\bibinfo {year} {1994})},\ \Eprint
  {http://arxiv.org/abs/arXiv:astro-ph/9312026} {arXiv:astro-ph/9312026}
  \BibitemShut {NoStop}%
\bibitem [{\citenamefont {{Padmanabhan}}(1996)}]{Padmanabhan1996}%
  \BibitemOpen
  \bibfield  {author} {\bibinfo {author} {\bibfnamefont {T.}~\bibnamefont
  {{Padmanabhan}}},\ }\href@noop {} {\emph {\bibinfo {title} {Current Applied
  Physics}}},\ edited by\ \bibinfo {editor} {\bibfnamefont {T.}~\bibnamefont
  {Padmanabhan}}\ (\bibinfo {year} {1996})\BibitemShut {NoStop}%
\bibitem [{\citenamefont {{Ohta}}\ \emph {et~al.}(2003)\citenamefont {{Ohta}},
  \citenamefont {{Kayo}},\ and\ \citenamefont {{Taruya}}}]{Ohta2003}%
  \BibitemOpen
  \bibfield  {author} {\bibinfo {author} {\bibfnamefont {Y.}~\bibnamefont
  {{Ohta}}}, \bibinfo {author} {\bibfnamefont {I.}~\bibnamefont {{Kayo}}}, \
  and\ \bibinfo {author} {\bibfnamefont {A.}~\bibnamefont {{Taruya}}},\ }\href
  {\doibase 10.1086/374375} {\bibfield  {journal} {\bibinfo  {journal} {\apj}\
  }\textbf {\bibinfo {volume} {589}},\ \bibinfo {pages} {1} (\bibinfo {year}
  {2003})},\ \Eprint {http://arxiv.org/abs/arXiv:astro-ph/0301567}
  {arXiv:astro-ph/0301567} \BibitemShut {NoStop}%
\bibitem [{\citenamefont {{Ohta}}\ \emph {et~al.}(2004)\citenamefont {{Ohta}},
  \citenamefont {{Kayo}},\ and\ \citenamefont {{Taruya}}}]{Ohta2004}%
  \BibitemOpen
  \bibfield  {author} {\bibinfo {author} {\bibfnamefont {Y.}~\bibnamefont
  {{Ohta}}}, \bibinfo {author} {\bibfnamefont {I.}~\bibnamefont {{Kayo}}}, \
  and\ \bibinfo {author} {\bibfnamefont {A.}~\bibnamefont {{Taruya}}},\ }\href
  {\doibase 10.1086/420762} {\bibfield  {journal} {\bibinfo  {journal} {\apj}\
  }\textbf {\bibinfo {volume} {608}},\ \bibinfo {pages} {647} (\bibinfo {year}
  {2004})},\ \Eprint {http://arxiv.org/abs/arXiv:astro-ph/0402618}
  {arXiv:astro-ph/0402618} \BibitemShut {NoStop}%
\bibitem [{\citenamefont {{Pace}}\ \emph {et~al.}(2010)\citenamefont {{Pace}},
  \citenamefont {{Waizmann}},\ and\ \citenamefont {{Bartelmann}}}]{Pace2010}%
  \BibitemOpen
  \bibfield  {author} {\bibinfo {author} {\bibfnamefont {F.}~\bibnamefont
  {{Pace}}}, \bibinfo {author} {\bibfnamefont {J.-C.}\ \bibnamefont
  {{Waizmann}}}, \ and\ \bibinfo {author} {\bibfnamefont {M.}~\bibnamefont
  {{Bartelmann}}},\ }\href {\doibase 10.1111/j.1365-2966.2010.16841.x}
  {\bibfield  {journal} {\bibinfo  {journal} {\mnras}\ }\textbf {\bibinfo
  {volume} {406}},\ \bibinfo {pages} {1865} (\bibinfo {year} {2010})},\ \Eprint
  {http://arxiv.org/abs/1005.0233} {arXiv:1005.0233} \BibitemShut {NoStop}%
\bibitem [{\citenamefont {{Lima}}\ \emph {et~al.}(1997)\citenamefont {{Lima}},
  \citenamefont {{Zanchin}},\ and\ \citenamefont {{Brandenberger}}}]{Lima1997}%
  \BibitemOpen
  \bibfield  {author} {\bibinfo {author} {\bibfnamefont {J.~A.~S.}\
  \bibnamefont {{Lima}}}, \bibinfo {author} {\bibfnamefont {V.}~\bibnamefont
  {{Zanchin}}}, \ and\ \bibinfo {author} {\bibfnamefont {R.}~\bibnamefont
  {{Brandenberger}}},\ }\href
  {http://articles.adsabs.harvard.edu/cgi-bin/nph-iarticle_query?1997MNRAS.291L...1L&amp;data_type=PDF_HIGH&amp;whole_paper=YES&amp;type=PRINTER&amp;filetype=.pdf}
  {\bibfield  {journal} {\bibinfo  {journal} {\mnras}\ }\textbf {\bibinfo
  {volume} {291}},\ \bibinfo {pages} {L1} (\bibinfo {year} {1997})},\ \Eprint
  {http://arxiv.org/abs/arXiv:astro-ph/9612166} {arXiv:astro-ph/9612166}
  \BibitemShut {NoStop}%
\bibitem [{\citenamefont {{Fosalba}}\ and\ \citenamefont
  {{Gazta\"naga}}(1998)}]{Fosalba1998a}%
  \BibitemOpen
  \bibfield  {author} {\bibinfo {author} {\bibfnamefont {P.}~\bibnamefont
  {{Fosalba}}}\ and\ \bibinfo {author} {\bibfnamefont {E.}~\bibnamefont
  {{Gazta\"naga}}},\ }\href {\doibase 10.1046/j.1365-8711.1998.02033.x}
  {\bibfield  {journal} {\bibinfo  {journal} {\mnras}\ }\textbf {\bibinfo
  {volume} {301}},\ \bibinfo {pages} {503} (\bibinfo {year} {1998})},\ \Eprint
  {http://arxiv.org/abs/arXiv:astro-ph/9712095} {arXiv:astro-ph/9712095}
  \BibitemShut {NoStop}%
\bibitem [{\citenamefont {{Engineer}}\ \emph {et~al.}(2000)\citenamefont
  {{Engineer}}, \citenamefont {{Kanekar}},\ and\ \citenamefont
  {{Padmanabhan}}}]{Engineer2000}%
  \BibitemOpen
  \bibfield  {author} {\bibinfo {author} {\bibfnamefont {S.}~\bibnamefont
  {{Engineer}}}, \bibinfo {author} {\bibfnamefont {N.}~\bibnamefont
  {{Kanekar}}}, \ and\ \bibinfo {author} {\bibfnamefont {T.}~\bibnamefont
  {{Padmanabhan}}},\ }\href {\doibase 10.1046/j.1365-8711.2000.03275.x}
  {\bibfield  {journal} {\bibinfo  {journal} {\mnras}\ }\textbf {\bibinfo
  {volume} {314}},\ \bibinfo {pages} {279} (\bibinfo {year} {2000})},\ \Eprint
  {http://arxiv.org/abs/astro-ph/9812452} {astro-ph/9812452} \BibitemShut
  {NoStop}%
\bibitem [{\citenamefont {{Peebles}}(1993)}]{Peebles1993}%
  \BibitemOpen
  \bibfield  {author} {\bibinfo {author} {\bibfnamefont {P.~J.~E.}\
  \bibnamefont {{Peebles}}},\ }\href@noop {} {\emph {\bibinfo {title}
  {Princeton Series in Physics, Princeton, NJ: Princeton University Press,
  |c1993}}},\ edited by\ \bibinfo {editor} {\bibfnamefont {P.~J.~E.}\
  \bibnamefont {Peebles}}\ (\bibinfo {year} {1993})\BibitemShut {NoStop}%
\bibitem [{\citenamefont {{Chernin}}(1993)}]{Chernin1993}%
  \BibitemOpen
  \bibfield  {author} {\bibinfo {author} {\bibfnamefont {A.~D.}\ \bibnamefont
  {{Chernin}}},\ }\href@noop {} {\bibfield  {journal} {\bibinfo  {journal}
  {\aap}\ }\textbf {\bibinfo {volume} {267}},\ \bibinfo {pages} {315} (\bibinfo
  {year} {1993})}\BibitemShut {NoStop}%
\bibitem [{\citenamefont {{Kashlinsky}}(1986)}]{Kashlinsky1986}%
  \BibitemOpen
  \bibfield  {author} {\bibinfo {author} {\bibfnamefont {A.}~\bibnamefont
  {{Kashlinsky}}},\ }\href {\doibase 10.1086/164350} {\bibfield  {journal}
  {\bibinfo  {journal} {\apj}\ }\textbf {\bibinfo {volume} {306}},\ \bibinfo
  {pages} {374} (\bibinfo {year} {1986})}\BibitemShut {NoStop}%
\bibitem [{\citenamefont {{Kashlinsky}}(1987)}]{Kashlinsky1987}%
  \BibitemOpen
  \bibfield  {author} {\bibinfo {author} {\bibfnamefont {A.}~\bibnamefont
  {{Kashlinsky}}},\ }\href {\doibase 10.1086/164895} {\bibfield  {journal}
  {\bibinfo  {journal} {\apj}\ }\textbf {\bibinfo {volume} {312}},\ \bibinfo
  {pages} {497} (\bibinfo {year} {1987})}\BibitemShut {NoStop}%
\bibitem [{\citenamefont {{Lahav}}\ \emph {et~al.}(1991)\citenamefont
  {{Lahav}}, \citenamefont {{Lilje}}, \citenamefont {{Primack}},\ and\
  \citenamefont {{Rees}}}]{Lahav1991}%
  \BibitemOpen
  \bibfield  {author} {\bibinfo {author} {\bibfnamefont {O.}~\bibnamefont
  {{Lahav}}}, \bibinfo {author} {\bibfnamefont {P.~B.}\ \bibnamefont
  {{Lilje}}}, \bibinfo {author} {\bibfnamefont {J.~R.}\ \bibnamefont
  {{Primack}}}, \ and\ \bibinfo {author} {\bibfnamefont {M.~J.}\ \bibnamefont
  {{Rees}}},\ }\href@noop {} {\bibfield  {journal} {\bibinfo  {journal}
  {\mnras}\ }\textbf {\bibinfo {volume} {251}},\ \bibinfo {pages} {128}
  (\bibinfo {year} {1991})}\BibitemShut {NoStop}%
\bibitem [{\citenamefont {{Bartlett}}\ and\ \citenamefont
  {{Silk}}(1993)}]{Bartlett1993}%
  \BibitemOpen
  \bibfield  {author} {\bibinfo {author} {\bibfnamefont {J.~G.}\ \bibnamefont
  {{Bartlett}}}\ and\ \bibinfo {author} {\bibfnamefont {J.}~\bibnamefont
  {{Silk}}},\ }\href {\doibase 10.1086/186802} {\bibfield  {journal} {\bibinfo
  {journal} {\apjl}\ }\textbf {\bibinfo {volume} {407}},\ \bibinfo {pages}
  {L45} (\bibinfo {year} {1993})}\BibitemShut {NoStop}%
\bibitem [{\citenamefont {{Del Popolo}}\ and\ \citenamefont
  {{Gambera}}(1998)}]{DelPopolo1998}%
  \BibitemOpen
  \bibfield  {author} {\bibinfo {author} {\bibfnamefont {A.}~\bibnamefont {{Del
  Popolo}}}\ and\ \bibinfo {author} {\bibfnamefont {M.}~\bibnamefont
  {{Gambera}}},\ }\href@noop {} {\bibfield  {journal} {\bibinfo  {journal}
  {\aap}\ }\textbf {\bibinfo {volume} {337}},\ \bibinfo {pages} {96} (\bibinfo
  {year} {1998})},\ \Eprint {http://arxiv.org/abs/astro-ph/9802214}
  {astro-ph/9802214} \BibitemShut {NoStop}%
\bibitem [{\citenamefont {{Del Popolo}}\ \emph {et~al.}(1998)\citenamefont
  {{Del Popolo}}, \citenamefont {{Gambera}},\ and\ \citenamefont
  {{Antonuccio-Delogu}}}]{DelPopolo1998a}%
  \BibitemOpen
  \bibfield  {author} {\bibinfo {author} {\bibfnamefont {A.}~\bibnamefont {{Del
  Popolo}}}, \bibinfo {author} {\bibfnamefont {M.}~\bibnamefont {{Gambera}}}, \
  and\ \bibinfo {author} {\bibfnamefont {V.}~\bibnamefont
  {{Antonuccio-Delogu}}},\ }\href {\doibase 10.1080/10556799808208151}
  {\bibfield  {journal} {\bibinfo  {journal} {Astronomical and Astrophysical
  Transactions}\ }\textbf {\bibinfo {volume} {16}},\ \bibinfo {pages} {127}
  (\bibinfo {year} {1998})}\BibitemShut {NoStop}%
\bibitem [{\citenamefont {{Del Popolo}}(2006)}]{DelPopolo2006b}%
  \BibitemOpen
  \bibfield  {author} {\bibinfo {author} {\bibfnamefont {A.}~\bibnamefont {{Del
  Popolo}}},\ }\href {\doibase 10.1051/0004-6361:20054441} {\bibfield
  {journal} {\bibinfo  {journal} {\aap}\ }\textbf {\bibinfo {volume} {454}},\
  \bibinfo {pages} {17} (\bibinfo {year} {2006})},\ \Eprint
  {http://arxiv.org/abs/0801.1086} {arXiv:0801.1086} \BibitemShut {NoStop}%
\bibitem [{\citenamefont {{Del Popolo}}\ \emph {et~al.}(2019)\citenamefont
  {{Del Popolo}}, \citenamefont {{Pace}},\ and\ \citenamefont
  {{Mota}}}]{DelPopolo2019}%
  \BibitemOpen
  \bibfield  {author} {\bibinfo {author} {\bibfnamefont {A.}~\bibnamefont {{Del
  Popolo}}}, \bibinfo {author} {\bibfnamefont {F.}~\bibnamefont {{Pace}}}, \
  and\ \bibinfo {author} {\bibfnamefont {D.~F.}\ \bibnamefont {{Mota}}},\
  }\href {\doibase 10.1103/PhysRevD.100.024013} {\bibfield  {journal} {\bibinfo
   {journal} {\prd}\ }\textbf {\bibinfo {volume} {100}},\ \bibinfo {pages}
  {024013} (\bibinfo {year} {2019})},\ \Eprint
  {http://arxiv.org/abs/1908.07322} {arXiv:1908.07322} \BibitemShut {NoStop}%
\bibitem [{\citenamefont {{Del Popolo}}\ \emph {et~al.}(2017)\citenamefont
  {{Del Popolo}}, \citenamefont {{Pace}},\ and\ \citenamefont {{Le
  Delliou}}}]{DelPopolo2017}%
  \BibitemOpen
  \bibfield  {author} {\bibinfo {author} {\bibfnamefont {A.}~\bibnamefont {{Del
  Popolo}}}, \bibinfo {author} {\bibfnamefont {F.}~\bibnamefont {{Pace}}}, \
  and\ \bibinfo {author} {\bibfnamefont {M.}~\bibnamefont {{Le Delliou}}},\
  }\href {\doibase 10.1088/1475-7516/2017/03/032} {\bibfield  {journal}
  {\bibinfo  {journal} {\jcap}\ }\textbf {\bibinfo {volume} {3}},\ \bibinfo
  {pages} {032} (\bibinfo {year} {2017})},\ \Eprint
  {http://arxiv.org/abs/1703.06918} {arXiv:1703.06918} \BibitemShut {NoStop}%
\bibitem [{\citenamefont {{Sheth}}\ \emph {et~al.}(2001)\citenamefont
  {{Sheth}}, \citenamefont {{Mo}},\ and\ \citenamefont {{Tormen}}}]{Sheth2001}%
  \BibitemOpen
  \bibfield  {author} {\bibinfo {author} {\bibfnamefont {R.~K.}\ \bibnamefont
  {{Sheth}}}, \bibinfo {author} {\bibfnamefont {H.~J.}\ \bibnamefont {{Mo}}}, \
  and\ \bibinfo {author} {\bibfnamefont {G.}~\bibnamefont {{Tormen}}},\ }\href
  {\doibase 10.1046/j.1365-8711.2001.04006.x} {\bibfield  {journal} {\bibinfo
  {journal} {\mnras}\ }\textbf {\bibinfo {volume} {323}},\ \bibinfo {pages} {1}
  (\bibinfo {year} {2001})},\ \Eprint
  {http://arxiv.org/abs/arXiv:astro-ph/9907024} {arXiv:astro-ph/9907024}
  \BibitemShut {NoStop}%
\bibitem [{\citenamefont {{Del Popolo}}\ and\ \citenamefont
  {{Gambera}}(2000)}]{DelPopolo2000}%
  \BibitemOpen
  \bibfield  {author} {\bibinfo {author} {\bibfnamefont {A.}~\bibnamefont {{Del
  Popolo}}}\ and\ \bibinfo {author} {\bibfnamefont {M.}~\bibnamefont
  {{Gambera}}},\ }\href@noop {} {\bibfield  {journal} {\bibinfo  {journal}
  {\aap}\ }\textbf {\bibinfo {volume} {357}},\ \bibinfo {pages} {809} (\bibinfo
  {year} {2000})},\ \Eprint {http://arxiv.org/abs/astro-ph/9909156}
  {astro-ph/9909156} \BibitemShut {NoStop}%
\bibitem [{\citenamefont {{Del Popolo}}\ \emph {et~al.}(2001)\citenamefont
  {{Del Popolo}}, \citenamefont {{Ercan}},\ and\ \citenamefont
  {{Xia}}}]{DelPopolo2001}%
  \BibitemOpen
  \bibfield  {author} {\bibinfo {author} {\bibfnamefont {A.}~\bibnamefont {{Del
  Popolo}}}, \bibinfo {author} {\bibfnamefont {E.~N.}\ \bibnamefont {{Ercan}}},
  \ and\ \bibinfo {author} {\bibfnamefont {Z.}~\bibnamefont {{Xia}}},\ }\href
  {\doibase 10.1086/321137} {\bibfield  {journal} {\bibinfo  {journal} {\aj}\
  }\textbf {\bibinfo {volume} {122}},\ \bibinfo {pages} {487} (\bibinfo {year}
  {2001})},\ \Eprint {http://arxiv.org/abs/astro-ph/0108080} {astro-ph/0108080}
  \BibitemShut {NoStop}%
\bibitem [{\citenamefont {{Del Popolo}}(2002{\natexlab{a}})}]{DelPopolo2002}%
  \BibitemOpen
  \bibfield  {author} {\bibinfo {author} {\bibfnamefont {A.}~\bibnamefont {{Del
  Popolo}}},\ }\href {\doibase 10.1051/0004-6361:20020399} {\bibfield
  {journal} {\bibinfo  {journal} {\aap}\ }\textbf {\bibinfo {volume} {387}},\
  \bibinfo {pages} {759} (\bibinfo {year} {2002}{\natexlab{a}})},\ \Eprint
  {http://arxiv.org/abs/astro-ph/0202436} {astro-ph/0202436} \BibitemShut
  {NoStop}%
\bibitem [{\citenamefont {{Peebles}}(1990)}]{Peebles1990}%
  \BibitemOpen
  \bibfield  {author} {\bibinfo {author} {\bibfnamefont {P.~J.~E.}\
  \bibnamefont {{Peebles}}},\ }\href {\doibase 10.1086/169456} {\bibfield
  {journal} {\bibinfo  {journal} {\apj}\ }\textbf {\bibinfo {volume} {365}},\
  \bibinfo {pages} {27} (\bibinfo {year} {1990})}\BibitemShut {NoStop}%
\bibitem [{\citenamefont {{Audit}}\ \emph {et~al.}(1997)\citenamefont
  {{Audit}}, \citenamefont {{Teyssier}},\ and\ \citenamefont
  {{Alimi}}}]{Audit1997}%
  \BibitemOpen
  \bibfield  {author} {\bibinfo {author} {\bibfnamefont {E.}~\bibnamefont
  {{Audit}}}, \bibinfo {author} {\bibfnamefont {R.}~\bibnamefont {{Teyssier}}},
  \ and\ \bibinfo {author} {\bibfnamefont {J.-M.}\ \bibnamefont {{Alimi}}},\
  }\href@noop {} {\bibfield  {journal} {\bibinfo  {journal} {\aap}\ }\textbf
  {\bibinfo {volume} {325}},\ \bibinfo {pages} {439} (\bibinfo {year}
  {1997})},\ \Eprint {http://arxiv.org/abs/astro-ph/9704023} {astro-ph/9704023}
  \BibitemShut {NoStop}%
\bibitem [{\citenamefont {{Del Popolo}}\ \emph {et~al.}(2005)\citenamefont
  {{Del Popolo}}, \citenamefont {{Hiotelis}},\ and\ \citenamefont
  {{Pe{\~n}arrubia}}}]{DelPopolo2005}%
  \BibitemOpen
  \bibfield  {author} {\bibinfo {author} {\bibfnamefont {A.}~\bibnamefont {{Del
  Popolo}}}, \bibinfo {author} {\bibfnamefont {N.}~\bibnamefont {{Hiotelis}}},
  \ and\ \bibinfo {author} {\bibfnamefont {J.}~\bibnamefont
  {{Pe{\~n}arrubia}}},\ }\href {\doibase 10.1086/429859} {\bibfield  {journal}
  {\bibinfo  {journal} {\apj}\ }\textbf {\bibinfo {volume} {628}},\ \bibinfo
  {pages} {76} (\bibinfo {year} {2005})},\ \Eprint
  {http://arxiv.org/abs/astro-ph/0508596} {arXiv:astro-ph/0508596} \BibitemShut
  {NoStop}%
\bibitem [{\citenamefont {{Del Popolo}}(2002{\natexlab{b}})}]{DelPopolo2002a}%
  \BibitemOpen
  \bibfield  {author} {\bibinfo {author} {\bibfnamefont {A.}~\bibnamefont {{Del
  Popolo}}},\ }\href {\doibase 10.1046/j.1365-8711.2002.05697.x} {\bibfield
  {journal} {\bibinfo  {journal} {\mnras}\ }\textbf {\bibinfo {volume} {336}},\
  \bibinfo {pages} {81} (\bibinfo {year} {2002}{\natexlab{b}})},\ \Eprint
  {http://arxiv.org/abs/astro-ph/0205449} {astro-ph/0205449} \BibitemShut
  {NoStop}%
\bibitem [{\citenamefont {{Albrecht}}\ and\ \citenamefont
  {{Skordis}}(2000)}]{Albrecht2000}%
  \BibitemOpen
  \bibfield  {author} {\bibinfo {author} {\bibfnamefont {A.}~\bibnamefont
  {{Albrecht}}}\ and\ \bibinfo {author} {\bibfnamefont {C.}~\bibnamefont
  {{Skordis}}},\ }\href {\doibase 10.1103/PhysRevLett.84.2076} {\bibfield
  {journal} {\bibinfo  {journal} {Physical Review Letters}\ }\textbf {\bibinfo
  {volume} {84}},\ \bibinfo {pages} {2076} (\bibinfo {year} {2000})},\ \Eprint
  {http://arxiv.org/abs/arXiv:astro-ph/9908085} {arXiv:astro-ph/9908085}
  \BibitemShut {NoStop}%
\bibitem [{\citenamefont {{Polisensky}}\ and\ \citenamefont
  {{Ricotti}}(2015)}]{Polisensky2015}%
  \BibitemOpen
  \bibfield  {author} {\bibinfo {author} {\bibfnamefont {E.}~\bibnamefont
  {{Polisensky}}}\ and\ \bibinfo {author} {\bibfnamefont {M.}~\bibnamefont
  {{Ricotti}}},\ }\href {\doibase 10.1093/mnras/stv736} {\bibfield  {journal}
  {\bibinfo  {journal} {\mnras}\ }\textbf {\bibinfo {volume} {450}},\ \bibinfo
  {pages} {2172} (\bibinfo {year} {2015})},\ \Eprint
  {http://arxiv.org/abs/1504.02126} {arXiv:1504.02126} \BibitemShut {NoStop}%
\bibitem [{\citenamefont {{Ryden}}(1988)}]{Ryden1988}%
  \BibitemOpen
  \bibfield  {author} {\bibinfo {author} {\bibfnamefont {B.~S.}\ \bibnamefont
  {{Ryden}}},\ }\href {\doibase 10.1086/166406} {\bibfield  {journal} {\bibinfo
   {journal} {\apj}\ }\textbf {\bibinfo {volume} {329}},\ \bibinfo {pages}
  {589} (\bibinfo {year} {1988})}\BibitemShut {NoStop}%
\bibitem [{\citenamefont {{Peirani}}\ and\ \citenamefont {{de Freitas
  Pacheco}}(2008)}]{Peirani2008}%
  \BibitemOpen
  \bibfield  {author} {\bibinfo {author} {\bibfnamefont {S.}~\bibnamefont
  {{Peirani}}}\ and\ \bibinfo {author} {\bibfnamefont {J.~A.}\ \bibnamefont
  {{de Freitas Pacheco}}},\ }\href {\doibase 10.1051/0004-6361:200809711}
  {\bibfield  {journal} {\bibinfo  {journal} {\aap}\ }\textbf {\bibinfo
  {volume} {488}},\ \bibinfo {pages} {845} (\bibinfo {year} {2008})},\ \Eprint
  {http://arxiv.org/abs/0806.4245} {arXiv:0806.4245} \BibitemShut {NoStop}%
\bibitem [{\citenamefont {{Karachentsev}}\ \emph {et~al.}(2002)\citenamefont
  {{Karachentsev}}, \citenamefont {{Dolphin}}, \citenamefont {{Geisler}},
  \citenamefont {{Grebel}}, \citenamefont {{Guhathakurta}}, \citenamefont
  {{Hodge}}, \citenamefont {{Karachentseva}}, \citenamefont {{Sarajedini}},
  \citenamefont {{Seitzer}},\ and\ \citenamefont
  {{Sharina}}}]{Karachentsev2002}%
  \BibitemOpen
  \bibfield  {author} {\bibinfo {author} {\bibfnamefont {I.~D.}\ \bibnamefont
  {{Karachentsev}}}, \bibinfo {author} {\bibfnamefont {A.~E.}\ \bibnamefont
  {{Dolphin}}}, \bibinfo {author} {\bibfnamefont {D.}~\bibnamefont
  {{Geisler}}}, \bibinfo {author} {\bibfnamefont {E.~K.}\ \bibnamefont
  {{Grebel}}}, \bibinfo {author} {\bibfnamefont {P.}~\bibnamefont
  {{Guhathakurta}}}, \bibinfo {author} {\bibfnamefont {P.~W.}\ \bibnamefont
  {{Hodge}}}, \bibinfo {author} {\bibfnamefont {V.~E.}\ \bibnamefont
  {{Karachentseva}}}, \bibinfo {author} {\bibfnamefont {A.}~\bibnamefont
  {{Sarajedini}}}, \bibinfo {author} {\bibfnamefont {P.}~\bibnamefont
  {{Seitzer}}}, \ and\ \bibinfo {author} {\bibfnamefont {M.~E.}\ \bibnamefont
  {{Sharina}}},\ }\href {\doibase 10.1051/0004-6361:20011741} {\bibfield
  {journal} {\bibinfo  {journal} {\aap}\ }\textbf {\bibinfo {volume} {383}},\
  \bibinfo {pages} {125} (\bibinfo {year} {2002})}\BibitemShut {NoStop}%
\bibitem [{\citenamefont {{Karachentsev}}(2005)}]{Karachentsev2005}%
  \BibitemOpen
  \bibfield  {author} {\bibinfo {author} {\bibfnamefont {I.~D.}\ \bibnamefont
  {{Karachentsev}}},\ }\href {\doibase 10.1086/426368} {\bibfield  {journal}
  {\bibinfo  {journal} {\aj}\ }\textbf {\bibinfo {volume} {129}},\ \bibinfo
  {pages} {178} (\bibinfo {year} {2005})},\ \Eprint
  {http://arxiv.org/abs/astro-ph/0410065} {arXiv:astro-ph/0410065} \BibitemShut
  {NoStop}%
\bibitem [{\citenamefont {{Peirani}}\ and\ \citenamefont {{de Freitas
  Pacheco}}(2006)}]{Peirani2006}%
  \BibitemOpen
  \bibfield  {author} {\bibinfo {author} {\bibfnamefont {S.}~\bibnamefont
  {{Peirani}}}\ and\ \bibinfo {author} {\bibfnamefont {J.~A.}\ \bibnamefont
  {{de Freitas Pacheco}}},\ }\href {\doibase 10.1016/j.newast.2005.08.008}
  {\bibfield  {journal} {\bibinfo  {journal} {\na}\ }\textbf {\bibinfo {volume}
  {11}},\ \bibinfo {pages} {325} (\bibinfo {year} {2006})},\ \Eprint
  {http://arxiv.org/abs/astro-ph/0508614} {arXiv:astro-ph/0508614} \BibitemShut
  {NoStop}%
\bibitem [{\citenamefont {{Bullock}}\ \emph {et~al.}(2001)\citenamefont
  {{Bullock}}, \citenamefont {{Kolatt}}, \citenamefont {{Sigad}}, \citenamefont
  {{Somerville}}, \citenamefont {{Kravtsov}}, \citenamefont {{Klypin}},
  \citenamefont {{Primack}},\ and\ \citenamefont {{Dekel}}}]{Bullock2001}%
  \BibitemOpen
  \bibfield  {author} {\bibinfo {author} {\bibfnamefont {J.~S.}\ \bibnamefont
  {{Bullock}}}, \bibinfo {author} {\bibfnamefont {T.~S.}\ \bibnamefont
  {{Kolatt}}}, \bibinfo {author} {\bibfnamefont {Y.}~\bibnamefont {{Sigad}}},
  \bibinfo {author} {\bibfnamefont {R.~S.}\ \bibnamefont {{Somerville}}},
  \bibinfo {author} {\bibfnamefont {A.~V.}\ \bibnamefont {{Kravtsov}}},
  \bibinfo {author} {\bibfnamefont {A.~A.}\ \bibnamefont {{Klypin}}}, \bibinfo
  {author} {\bibfnamefont {J.~R.}\ \bibnamefont {{Primack}}}, \ and\ \bibinfo
  {author} {\bibfnamefont {A.}~\bibnamefont {{Dekel}}},\ }\href {\doibase
  10.1046/j.1365-8711.2001.04068.x} {\bibfield  {journal} {\bibinfo  {journal}
  {\mnras}\ }\textbf {\bibinfo {volume} {321}},\ \bibinfo {pages} {559}
  (\bibinfo {year} {2001})},\ \Eprint
  {http://arxiv.org/abs/arXiv:astro-ph/9908159} {arXiv:astro-ph/9908159}
  \BibitemShut {NoStop}%
\end{thebibliography}

%

\end{document}